\documentclass[10pt,letterpaper]{JHEP3}
\usepackage{amsmath}
\usepackage{amssymb}
\usepackage{amsfonts}
\usepackage{lscape, graphicx}
\usepackage{cite}       
\usepackage{bm}         

\usepackage[all]{xy}
\advance\parskip 1.0pt plus 1.0pt minus 2.0pt   
\def\href#1#2{#2}	

\def\coeff#1#2{{\textstyle {\frac {#1}{#2}}}}
\def\half{\coeff 12}

\def\N{{\cal N}}

\def\R{{\mathbb R}}

\def\tr{{\rm tr}}

\def\Z{{\mathbb Z}}

\def\Dslash{{\rlap{\raise 1pt \hbox{$\>/$}}D}}
\def\O{{\cal O}}

\preprint{SLAC-PUB-13666}

\title{ Conformality or  confinement:\\  (IR)relevance of topological excitations }

\author
    {
    {
    \def\href#1#2{#2}	
    Erich Poppitz$^1$\footnote{\email{poppitz@physics.utoronto.ca}}~
    and Mithat \"Unsal$^2$\footnote{\email{unsal@slac.stanford.edu}}~
           \\${}^1$Department of Physics, University of Toronto,
    Toronto, ON M5S 1A7, Canada
     \\${}^2$SLAC and Physics Department, Stanford University, Stanford, CA 94025/94305, USA
        }
    }%

\abstract{

\smallskip

{\small{

What distinguishes two asymptotically-free non-abelian  gauge theories  on ${\bf  R}^4$, one of which is just below  the conformal window boundary and confines,  while the other is slightly  above the  boundary and flows to an infrared conformal field theory?  
In this work, we aim to answer this question for  non-supersymmetric Yang-Mills theories with fermions in arbitrary chiral or vectorlike representations.  We use the presence or absence of mass gap for gauge fluctuations as an identifier of the infrared behavior.  
With the present-day understanding of such gauge theories, the  mass gap for gauge fluctuations cannot be computed   on  ${\bf  R}^4$. However, recent progress allows its  non-perturbative  computation  on $ {\bf R}^3 \times  {\bf S}^1$  by using    either the twisted partition function or deformation theory, for a range of sizes of  ${\bf S}^1$  depending on the theory. For small number of fermions, $N_f$, 
we show  that the mass gap increases with increasing radius, due to the   
 non-dilution  of  monopoles and  bions---the topological excitations relevant for confinement on $ {\bf R}^3 \times  {\bf S}^1$.
For sufficiently large  $N_f$, we show that the mass gap decreases with increasing radius.
In a class of theories, we claim  that the decompactification limit can be taken  while remaining within the region of validity of  semiclassical techniques, giving 
the first  examples of   semiclassically  solvable Yang-Mills theories  at  any size ${\bf S}^1$. 
For general non-supersymmetric vectorlike or chiral theories, we conjecture that  the change in the behavior of the mass gap on $ {\bf R}^3 \times  {\bf S}^1$ as a function of the radius 
occurs near the lower  boundary of the conformal window and 
 give non-perturbative estimates of  its value. For vectorlike theories, we compare our  estimates of the conformal window with existing lattice results,  truncations of the Schwinger-Dyson equations, NSVZ beta function-inspired estimates, and degree of freedom counting criteria. For  multi-generation chiral gauge theories, to the best of our knowledge, our estimates of the conformal window are the only  known ones. }
      }
      }

\begin{document}

\maketitle
 
 \section{Introduction and summary}

\subsection{Phenomenological motivation}
 
There exist numerous  suggestions that (near-)conformal  strong gauge dynamics can address various problems in models of elementary particle physics. Perhaps, the most pressing issue in particle physics,  to be  studied at the LHC,  is the problem of electroweak 
symmetry breaking. A strongly-coupled, vectorlike or chiral, gauge theory without  elementary 
Higgs scalars may in principle induce the dynamical breaking of electroweak symmetry. 
Using a scaled-up version of QCD to this end, however, is ruled out by electroweak 
precision data. In addition, in its minimal form, the model fails to produce a satisfactory spectrum of the standard model particles.  It has been argued that gauge sectors with  near-conformal or conformal behavior can help solve phenomenological problems of fine-tuning and flavor in models of dynamical electroweak symmetry breaking: see  \cite{Holdom:1981rm, Yamawaki:1985zg, Appelquist:1986an} for early references on ``walking" technicolor and the more recent  ``conformal technicolor" proposal of \cite{Luty:2004ye}. Conformal sectors  are   an integral part of the recent ``unparticle" paradigm  \cite{Georgi:2007ek}. There is also a vast literature on weak-scale model building using  warped extra dimensions, inspired by Randall-Sundrum models \cite{Randall:1999ee}. These models are thought to be dual to (broken) conformal field theories, see  \cite{ArkaniHamed:2000ds} and references therein.

However, our current understanding of strongly-coupled gauge theory dynamics is rather poor and 
   phenomenological models using near-conformal dynamics have to often rely on dynamical assumptions. Even the question of which asymptotically free gauge theories flow to conformal field theories   in the infrared is not satisfactorily answered, not to speak of  having control over properties such as the scaling dimensions of various operators, see 
\cite{Rattazzi:2008pe, Rychkov:2009ij} for recent first-principle studies. This an important theoretical issue, whose better understanding may be of future phenomenological relevance.

 \subsection{Statement of the confinement-conformality problem}

Consider   Yang-Mills (YM) theory  with massless vectorlike  or chiral fermions on ${\bf R}^4$.  
The gauge group is chosen to be  ${\rm SU}(N)$  and we assume that the chiral fermions transform as one- or two-index  representations of   ${\rm SU}(N)$. 
The perturbative renormalization group beta function can be used to determine whether such theories  are   asymptotically free or infrared free, as a function of the number of fermion representations, $N_f$. 
Gauge theories with $N_f >  N_f^{AF}$, where $N_f^{AF}$ indicates the boundary of asymptotic freedom, are  infrared free. Theories with $N_f <  N_f^{AF}$ are believed 
to exhibit  two types of behavior in the infrared (IR).  When the number of flavors is sufficiently small 
$(N_f < N^{*}_f)$,  it is believed that  confinement takes place.  If the number of flavors 
is in the  $N_f^{*} < N < N_f^{AF}$ range, it is believed that  the theory is in an interacting  non-abelian Coulomb phase  in the IR. These are IR  conformal field theories (CFT) and  the corresponding range of $N_f$ is   referred to as the   ``conformal window."  

For a subclass of QCD-like gauge theories with fundamental matter, 
when  $N_f$ is close to the upper boundary of the conformal window, the existence of a conformal fixed point can be shown reliably within perturbation theory \cite{Banks:1981nn}.   Such fixed points are called Banks-Zaks fixed  points and have a tunable small coupling. Theories with two-index representation fermions, including multi-generation chiral gauge theories, do not possess fixed points with tunably small coupling constants and thus have no  Banks-Zaks limit. However,  it is 
 believed that they have  conformal windows as well. The standard expectation  regarding the phases of such gauge theories can be shown in a simple phase diagram:\footnote{
$N_f$ 
is discrete. If we were to plot the same diagram as a function of $N_f/N_c$, for theories admitting the Banks-Zaks limit, one can replace it with a continuous variable.} 
 \begin{equation} 
 \xy 
 (-60,0)*{\bullet}; 
(-40,0)*{\bullet}; 
(-80,0)**\dir{-} ?(.85)+(3,3)*{\scriptstyle {\rm confined}}; 
(0,0)*{\; \; \;  N_f}; 
(-40,0)**\dir{-} ?(0.05)*\dir{<} ; 
(-20,3)*{\scriptstyle {\rm IR-free}} ; 
(-50,3)*{\scriptstyle {\rm IR-CFT}} ; 
(-6,2)*{\scriptstyle {\infty}} ; 
(-40,5)*{\scriptstyle  N_f^{AF}} ; 
(-60,5)*{\scriptstyle N_f^{*}} ; 
\endxy 
\label{phasediag} 
\end{equation}
There are multiple  (well-known)  conceptual questions regarding these theories: 
\begin{itemize}
{\item{\bf Mechanisms of confinement and conformality:} What distinguishes  two  theories, one   just below  the conformal boundary and confining,   the other slightly  above the conformal window boundary? In other words, why does a confining gauge theory confine and why does an IR-CFT, with an almost identical microscopic matter  content,  deconfine?}
{\item{\bf Type of phase transition:} What is the nature of phase transition from the confined phase to the IR-CFT phase, for example, in theories where $N_f/N_c$ can be continuously tuned? }
 {\item{\bf Lower boundary of conformal window:} What is the physics determining the boundary of conformal window?}
\end{itemize}

Many  of the questions  regarding non-supersymmetric vectorlike or chiral gauge theories on ${\bf R}^4$  are  beyond the scope of our current analytical understanding of quantum Yang-Mills theories  and we will not come to grips with them fully.  At this front, one may  hope that numerical lattice gauge theory may be of  use. However, there are  well-known practical   difficulties in lattice gauge theories with vectorlike fermions in the chiral limit,  and much more severe   difficulties  
for theories with chiral matter.   Furthermore, a practical  lattice gauge theory simulation is set  on  ${\bf T}^4$.  For confining gauge theories, the notion of  sufficiently large  ${\bf T}^4$ is meaningful and provides  a good description  of 
the target theory on ${\bf R}^4$.  On the other hand,  if the target theory is conformal,  the analysis of the  lattice theory  will be more tricky due to finite volume effects.   

\subsection{Review of known diagnostics of confinement vs. conformality}

We begin by  reviewing the known approaches to determining the conformal window in nonsupersymmetric theories.
As will become clear from our discussion below, we believe that, up-to-date, only the lattice approach offers a controllable first-principle determination. However, the lattice still suffers from technical difficulties (for general gauge groups and representations) and, in its current state, only works for vectorlike  theories. 
Thus, it is worthwhile to study new approaches to the conformality-confinement transition, applicable to any theory, as they might provide further insight and guide future studies. 

The main idea of this paper is  to approach the problem  by studying the 
behavior of the mass gap for gauge fluctuations as a function of the volume. We consider the gauge theories of interest in a centrally-symmetric vacuum on a partially compactified geometry ${\bf R}^3 \times {\bf S}^1$ and use the fact that at small  ${\bf S}^1$ the mass gap and its volume dependence can be reliably calculated. We find that the volume dependence of the mass gap changes as $N_f$ is increased and use this to estimate the boundary of conformal window. While our estimate of the conformal window in $\bf{R}^4$ is based  on a conjecture---a quality shared by most other estimates---our results  are similar (for vectorlike theories,   where a comparison can be made) to those of previous analytic approaches, see \S\ref{comparison}. This similarity occurs, somewhat to our surprise, despite the quite different framework used.\footnote{Our analysis invokes only semiclassical methods, the index theorem on ${\bf R}^3 \times {\bf S}^1$,  and the one-loop beta function, while all other approaches rely on (at least) the two-loop beta function.} For multiple-generation  or ``quiver" chiral gauge theories, see \S\ref{chiral}, our estimates of the conformal window are the only ones we are aware of.

\subsubsection{Truncated  Schwinger-Dyson equations}  
\label{SDeqns}
 
Most work in the literature is focused on vectorlike gauge theories and uses a  fermion bilinear condensate  $\langle \bar\psi \psi \rangle$ as an order parameter  to identify the conformal window. The basic idea  behind this approach is the standard assumption that 
confining non-supersymmetric vectorlike gauge theories  on ${\bf R}^4$ will also exhibit chiral symmetry breaking ($\chi$SB). An IR-conformal field theory, on the other hand,  is  free of any dynamically generated scale or any (chiral) condensates. Then, the  expected phase diagram is: 
 \begin{equation} 
 \xy 
 (-60,0)*{\bullet}; 
(-40,0)*{\bullet}; 
(-80,0)**\dir{-} ?(.85)+(3,3)*{\scriptstyle \langle \bar\psi \psi \rangle \neq 0 }; 
(0,0)*{\; \; \;  N_f}; 
(-40,0)**\dir{-} ?(0.05)*\dir{<} ; 
(-20,3)*{\scriptstyle {\rm IR-free}} ; 
(-50,3)*{\scriptstyle  \langle \bar\psi \psi \rangle =  0 } ; 
(-6,2)*{\scriptstyle {\infty}} ; 
(-40,5)*{\scriptstyle  N_f^{AF}} ; 
(-60,5)*{\scriptstyle N_f^{*}} ; 
\endxy 
\label{phasediag2} 
\end{equation}

This idea is usually implemented in the  ladder approximation to the Schwinger-Dyson equations for the chiral condensate 
(the ``gap equation," see \cite{Peskin:1982mu} for a clear and up-to-date introduction). The critical value of the coupling that triggers $\chi$SB corresponds to a 
large anomalous dimension of the fermion bilinear, typically $\gamma \approx 1 $. The critical coupling, if  reached at the (putative) fixed point $\alpha_*$ of the beta function,  triggers chiral symmetry breaking, generates a dynamical mass for the fermions, and changes the flow of the coupling away from the fixed point to that of the pure YM theory. In contrast, if the fixed-point coupling is smaller than the critical value, $\gamma(\alpha_*) < 1$, chiral symmetry is unbroken and the expected behavior is   IR-CFT. Comparing the critical and fixed-point couplings as a function of $N_f$ allows for an estimate of $N_f^*$, see e.g.~\cite{Appelquist:1988yc, Cohen:1988sq, Miransky:1996pd, Appelquist:1996dq, Appelquist:1998rb,Dietrich:2006cm}

The validity of the approximations used in deriving the bounds on the conformal window in the Schwinger-Dyson approach is  discussed in the literature (see \cite{Appelquist:1988yc}, as well as \cite{Gies:2005as}, which uses another formalism combined with higher-order calculations,  to obtain somewhat different, typically lower, estimates of the lower boundary, $N_f^*$, of the conformal window in vectorlike theories). Usually, the precision of results from  the  Schwinger-Dyson equations is gauged by estimating their variation  due to the next-order loop correction. While this may be a useful guide, we note that the perturbative loop expansion misses nonperturbatively generated multi-fermion interactions  due to  nontrivial topological excitations (such as instantons, instanton molecules, or instanton ``quarks") that become important near the transition, where the coupling is typically strong, see \cite{Appelquist:1997dc, Velkovsky:1997fe}. Hence, the errors inherent to the Schwinger-Dyson approach are, most likely, underestimated. 

Leaving aside a precise estimate of the  uncertainty due to truncating and approximating  the Schwinger-Dyson equations  (as well as the critical  and fixed-point values of the coupling), 
we note that an approach to determining the conformal window using  equations  for the fermion 
propagator  is not appropriate in chiral gauge theories. In such theories, an expectation value of the fermion bilinear is forbidden by gauge invariance. Furthermore,    there are well-known examples of chiral gauge theories believed to exhibit {\it confinement without  } $\chi$SB, the classic example \cite{Dimopoulos:1980hn} being an SU$(5)$ gauge theory with ${\bf 5}$ and a ${\overline {\bf 10}}$ representation Weyl fermions. Chiral gauge theories with a large number of  generations or with added extra vectorlike matter, however, are also expected to have conformal windows, hence it is desirable to develop techniques that are also applicable to such theories.

\subsubsection{Supersymmetry-inspired estimates}

In ${\cal{N}} = 1$ supersymmetric theories,
there is  a     relation between the anomalous dimensions of  chiral matter superfields   and the all-order NSVZ beta-function of the gauge coupling. In addition, the superconformal algebra relates the dimensions of  chiral operators at an IR fixed point to their $R$-charge. Together, these properties allow for a determination of  the boundary of the conformal window in many supersymmetric examples, which passes many nontrivial tests (see \cite{Intriligator:1995au} for a review and references).\footnote{ 
In this paper, we will not discuss  supersymmetric theories and their IR fixed points, where the analysis  is complicated by the presence of scalars and quantum moduli spaces. We only note that 
other criteria \cite{Intriligator:2005if} have been conjectured to distinguish conformality and confinement in more involved, e.g.~chiral, supersymmetric cases.}

 A similar, but, unlike NSVZ, not derived\footnote{Recall the simple reasoning  \cite{Novikov:1983uc, Novikov:1985rd} leading to the NSVZ beta function: the  bosonic and fermionic zero modes  in an instanton background give rise to the one-loop running of the gauge coupling entering the instanton vertex, while the nonzero modes' contributions cancel to all orders of perturbation theory due to supersymmetry. Thus, in ${\cal N} = 1$ supersymmetry, the only  higher-loop renormalization of the instanton vertex  is from the $Z$-factors of the zero modes, hence the NSVZ relation between anomalous dimensions and the beta function of the coupling appears quite naturally. Clearly, a similar route to  establish an NSVZ-like formula does not apply without SUSY, hence its conjectural nature.} by combining instanton calculus  with supersymmetric  nonrenormalization theorems, relation between the anomalous dimensions of fermion matter fields and the all-order beta function was conjectured  for nonsupersymmetric theories in \cite{Ryttov:2007cx}. It was noted there that it implies  features qualitatively similar to the NSVZ formula: as $N_f$  decreases away from the asymptotic freedom boundary, the value of the anomalous dimension of the fermion bilinear at the zero of the beta function (the putative IR fixed point) increases. The criterion 
 $\gamma(\bar\psi \psi) \le 2$---the unitarity bound, see e.g.~\cite{Grinstein:2008qk}, on the fermion bilinear scalar operator---was used to place a bound on the lower end of the conformal window, $N_f^*$ \cite{Ryttov:2007sr}. 
Since  the validity of the NSVZ-inspired beta-function for nonsupersymmetric theories is unclear,  whether 
these estimates are upper or lower bounds on $N_f^*$ is uncertain. In addition, as in supersymmetry, the NSVZ-inspired estimates are more difficult (but not impossible) to apply to chiral gauge theories. 

Note also the work of ref.~\cite{Chishtie:1999tx} on the application of Pad\` e approximations to the beta function for estimates of the conformal window (the results are quoted in section \ref{sdcomparison}).

\subsubsection{Degree of freedom counting via thermal inequality}

A constraint  on the massless spectrum of   strongly interacting asymptotically free gauge theories was conjectured in \cite{Appelquist:1999hr}, 
based on the expectation that the number of degrees of freedom  decreases along the renormalization  flow. It was conjectured 
that in the zero-temperature limit, the free energy of the massless IR degrees of freedom 
should be smaller than or equal to the similar quantity calculated for the massless fundamental (UV) degrees of 
freedom. The free energy inequality offers a simple degree-of-freedom counting criterion, which can be applied 
 whenever a guess for the massless degrees of freedom of a strongly-coupled theory can be made (usually based on anomaly matching) and their free energy reliably computed.

   In vectorlike SU$(N)$ gauge theories with $N_f$ massless fundamental flavors, ref.~\cite{Appelquist:1999hr} found that it implies that  the conformal window  disappears for $N_f< 4N$  for $N$ sufficiently large.\footnote{See the discussion in  section V.A.5 of 
   \cite{Sannino:2005sk} of the special case of $N=2$, pointing out that application of the  thermal inequality, assuming breakdown of the enhanced SU$(2N_f)$ global symmetry to SP$(2N_f)$ in the confining phase, yields $N_f^* = 4.74$, in conflict with the Schwinger-Dyson equations estimate  $N_f^* = 7.86$.}
   For higher-dimensional (two-index) vectorlike representations, however, the inequality of \cite{Appelquist:1999hr} yields no constraints on the conformal window \cite{Sannino:2005sk}, and for the  multiple-generations chiral theories of \S\ref{chiral} the inequality has not been applied.

\subsubsection{Lattice gauge theory} 

The current state of the art allows  lattice simulations of vectorlike gauge theories only. 
There have been several recent studies of the confinement to conformality transition in vectorlike SU$(N)$, $N=2,3$, gauge 
theories with one- or two-index fermion representations. We give references and a summary of recent lattice results in \S\ref{lattice}.

\subsection{Mass gap for gauge fluctuations and  the  onset of  conformality}
It is desirable to  probe  confinement and conformality more directly, without any recourse to 
chiral symmetry realization and the fermion bilinear condensate, which does not always apply. 
Phases of gauge theories  
may  be classified by using  Wilson loop or   't Hooft loop (disorder) operators.   Below, we will give some motivational description in terms of the Wilson operator  $W(C)$  on  
$ {\bf R}^4$ for various gauge theories as a function of $N_f$.  In  theories without fundamental matter (or   matter which can screen arbitrary external charges),  the
Wilson loop can be used to deduce the long distance interaction between test charges.  Let 
$\Sigma \subset {\bf R}^4$ denote a large rectangular surface with boundary    $C =  \partial  \Sigma$ and  area is ${\rm Area}( \Sigma) = r \times T$. 
Then, 
 \begin{eqnarray}
&&\lim_{T \rightarrow \infty} \frac{1}{T} \log  \langle W(C) \rangle \sim  \left\{
\begin{array}{l l}
  \sigma r, &
 {\rm confined} \cr 
   \frac{g^2_*}{r}  &   {\rm  IR-CFT}
   \end{array} \right.
 \end{eqnarray}
where $\sim$ sign refers to the asymptotic nature of these formulas,  
 $\sigma$ is the string tension,  and $g_*$ is a coupling constant. 
 For theories with sufficiently few fundamental matter fields, the expectation value of large Wilson 
 loop is expected to obey  $\langle W(C) \rangle \sim c_1 e^{- \sigma {\rm Area}(\Sigma) }+ c_2  e^{-\mu {\rm Perimeter}(C)}$.

The most robust  aspect of the phase diagram given in (\ref{phasediag}) is that for confining  theories,      the gauge fluctuations are always gapped (short-ranged) and   for 
IR-CFTs,  the gauge fluctuations are always massless, hence infinite-range.  The natural scale  of gauge fluctuations is  the characteristic size of flux tubes for these gauge theories on 
$ {\bf R}^4$:
 \begin{eqnarray}
&& m^{-1}_{\rm gauge \; fluc.}  ({\bf R}^4)= 
  \left\{
\begin{array}{lll}
  {\rm finite}& \;\;\;  N_f < N_f^{*}  & \qquad {\rm confined}   \cr 
\infty  & \;\;\;   N_f^{*} <  N_f <  N_f^{AF}   &\qquad  {\rm  IR-CFT}
   \end{array} \right.
   \label{IRCCC}
 \end{eqnarray}
 We refer to this  characterization  of   IR  confinement-conformality 
as {\it mass gap for gauge fluctuations} criterion.   

We suggest the following simple picture regarding the transition from confined to conformal phase. Consider a class of gauge theories with    a fixed small coupling  $g^2(\mu)$
at the UV cutoff $\mu$.  For zero flavors, this is pure YM theory on ${\bf R}^4$.  This theory is believed to possess a mass gap   and exhibit confinement.    The mass gap is  of the order of the inverse of the flux tube thickness. 
 As the number of fermion representations $N_f$ is increased while remaining in the range  $N_f < N_f^{*}$ and   holding $g^2(\mu)$ fixed,  the range of the gauge fluctuations will increase gradually.  
  Upon crossing the conformal window boundary, for $ N_f^{*} <  N_f <  N_f^{AF} $, the screening effects of the fermions cause the gauge fluctuations to  become infinite range and the mass gap to vanish. 
  
Unfortunately, there is no known analytic way to quantify this picture on  $ {\bf R}^4$ in a reliable manner.  In particular, on ${\bf R}^4$, we do not know how to calculate the mass gap  (or absence thereof)  for gauge fluctuations  for a given gauge theory. 
Obviously, pursuing the criterion of eqn.~(\ref{IRCCC}) to the confining-conformality problem  does not {\it a priori}  strike  us as  a smart strategy, as it  maps  the problem on the onset 
of the conformal window to the ``mass gap" problem for gauge fluctuations. 

Recently, however,  there has  been significant progress in our understanding of  
non-supersymmetric  gauge theories   by using  circle compactification down to  $ {\bf R}^3 \times {\bf S}^1$ \cite{Unsal:2007jx, Shifman:2008ja, Unsal:2008ch, Shifman:2008cx, Shifman:2009tp, Poppitz:2009kz}.
Circle compactification was used earlier in the supersymmetric context  as a controllable deformation of supersymmetric gauge theories \cite{Seiberg:1996nz, Aharony:1997bx}. Many results regarding 
supersymmetric YM theories on  $ {\bf R}^4$ were obtained by starting from a finite-size $ {\bf R}^3 \times {\bf S}^1 $---such as the chiral condensate \cite{Davies:1999uw}, mass gap, and moduli spaces. In the non-supersymmetric case, the new ingredient is the use of the 
twisted partition function and deformation theory, which make the study of non-supersymmetric theories at small ${\bf S^1}$ as tractable as the supersymmetric ones.
  In some theories, it has been conjectured that the   small and large 
${\bf S}^1 $ regimes  are smoothly connected---the {\it ``smoothness conjecture"}---and in many cases, lattice studies have shown results supporting this conjecture \cite{Myers:2007vc, Ogilvie:2007tj, Myers:2008zm, Cossu:2009sq, Meisinger:2009ne}. A more refined version of the smoothness conjecture is at the heart of the study of our confinement-conformality diagnostic, as we  explain in the following sections.

\subsection{Circle compactification, twisted partition function, and  deformation theory}

In this section, we recall the main features of the compactification on a non-thermal circle (the twisted partition function) and of deformation theory which make center-symmetric theories on small $\bf{S}^1$ semiclassically solvable. 

 Circle compactification is a quantization of a gauge theory on  ${\bf R}^3 \times {\bf S}^1$ by using 
 translationally invariant {\it periodic} boundary conditions for fermions. Consequently, the  path integral  translates, in the operator formalism,  not to the thermal  partition function, but to a zero-temperature ``twisted" partition function:
\begin{equation}
\widetilde Z (L)= \tr \left[ e^{-L H} (-1)^F \right]
\label{twistedz}
\end{equation}
where $H$ is Hamiltonian, $ \tr $ is over the Hilbert space,   $L >0 $ is the circumference of $S^1$, and  $(-1)^F$ is fermion number modulo two.\footnote{ \label{pbc} The periodic boundary conditions are sometime referred to as ``unphysical" boundary conditions. 
This characterization is incorrect. $\widetilde Z (L)$ may be used to observe zero temperature quantum phase transitions as a function of compactification scale  $L$,  which are generically richer than thermal transitions. Moreover, in supersymmetric theories, $\widetilde Z (L)$ is the 
usual supersymmetric index and is  independent of ${\bf S}^1$ size. }   Unlike thermal compactification, where all asymptotically free confining gauge theories undergo a transition to a deconfined quark-gluon phase at sufficiently high temperature, in a circle compactification, there may not be any phase transitions at all.  

For example, in 
pure $\N=1$ supersymmetric YM (SYM) theories with supersymmetry preserving boundary conditions, there is compelling reason to believe that there is no phase transition as a function of radius. 
Recently, the first examples of non-supersymmetric gauge theories  which do not
 undergo a  center-symmetry changing  
(the analog of confinement-deconfinement)  phase transition were  also found. Such center-symmetric theories  provide new insights into gauge dynamics and shed light on the confinement and mass gap problem \cite{ Unsal:2007jx}. 
 Soon after these examples,   deformations of Yang-Mills  theories  and QCD  on ${\bf R}^3 \times {\bf S}^1$ that we refer to as YM$^*$ and QCD$^*$  were constructed in \cite{Shifman:2008ja, Unsal:2008ch}. The role of the 
 deformation is to stabilize center symmetry and guarantee that, at least in the sense of center symmetry, the small-${\bf S}^1$ YM$^*$ theory
 can be smoothly connected to 
 the corresponding YM theory at arbitrarily large ${\bf S}^1$. Furthermore, in the  center-symmetric small-${\bf S}^1$ YM$^*$ theory confinement can be understood quantitatively in the semiclassical approximation. 
  
   Currently, with the help of deformation theory, we have a  detailed and quantitative understanding of gauge theories on 
  ${\bf R}^3 \times {\bf S}^1$ at sufficiently small $L$, including many chiral gauge theories \cite{Shifman:2008cx, Poppitz:2009kz}.  Perhaps to the surprise of the past,  the topological excitations which lead  to a mass gap in almost all theories are not monopoles (or more precisely monopole-instantons), but rather more exotic---and not (anti-)self-dual---excitations. These are the magnetic ``bions", ``quintets", and other interesting composites of the fundamental monopoles and Kaluza-Klein 
  antimonopoles.    As already mentioned and further explained below, in center-symmetric theories at small $L$ we can reliably evaluate the mass gap for gauge fluctuations by using semiclassical techniques.  
  The mass gap depends on the details of the theory, 
  such as the rank of the gauge group, the number of flavors, and matter content (representations 
  ${\cal R}$ of multiplicity $N_f$):  
  \begin{equation}
   m^{-1}_{\rm gauge \; fluct.}  ({\bf R}^3 \times {\bf S}^1)= F (L, g(L), N, N_f, {\cal R})~.
   \label{fL}
  \end{equation}
  If we can take  the $L \rightarrow \infty $ limit of (\ref{fL}), we can also deduce, according to our mass gap  criterion  (\ref{IRCCC}),  which theory will confine and which will exhibit conformality.    However, in ref.~\cite{Shifman:2008cx} (and others), 
  we have typically considered few-flavor theories with a strong scale $\Lambda$. In 
   these theories, the condition of validity of semiclassical approximation is that the size of $\bf{S}^1$ is sufficiently small:\footnote{Strictly speaking, at arbitrary $N$,  this window is  $\Lambda  LN \ll 1$. In particular, at large $N$, the region of validity of any semiclassical analysis of the center-symmetric theory shrinks to zero in compliance with 
   large-$N$ volume independence. This also means that, at $N=\infty$, center-symmetric theories formulated on  ${\bf R}^{4-d} \times {\bf T}^d$  lack  a weak coupling description regardless of the size of  $d$-torus, ${\bf T}^d$, see \cite{Unsal:2008ch}.
} 
  \begin{equation}
\Lambda L  \ll 1 ~.
  \end{equation}
Thus, in confining gauge theories,  we can never take the decompactification limit while  remaining 
within the semiclassical window. This is of course, understandable.  The semiclassical approximation requires diluteness of topological excitations and for confining gauge theories when $\Lambda L \sim  1$  the topological excitations become non-dilute.

\subsection{Dilution of  topological excitations  on  ${\bf R}^3 \times {\bf S}^1$
and conformality on ${\bf R}^4$} 

In this section, we discuss  a calculable example of the main effect providing the intuition behind our estimate of the conformal window---we argue that in theories that become conformal on ${\bf R}^4$  we expect  the effects of   topological excitations to be diluted in the large volume limit.

To this end, we first construct an example of an asymptotically free non-supersymmetric gauge theory where one can take 
  the decompactification limit (or take arbitrarily large $L$) while remaining within theoretical control.    This gauge theory can thus be studied (even without lattice) by using semiclassical techniques and  bounds on the mass gap can be obtained, exhibiting absence or presence of  confinement.   
  Such theories, {\bf  a.)}~must be  close to the upper boundary of the conformal window with a weakly coupled infrared fixed point, and,  {\bf  b.)}~upon compactification  on ${\bf R}^3 \times {\bf S}^1$,   they  must  not break their center symmetry. Even if center symmetry breaks, there must exist a repulsion between the eigenvalues of the 
 Wilson line on ${\bf S}^1$ leading to dynamical abelianization without charged massless fermions. These two criteria guarantee  that the theory is weakly coupled on ${\bf R}^4$ ({\bf a.}) and   that at finite $L$ it abelianizes and remains  weakly-coupled, instead of flowing to  strong coupling  ({\bf b.}).

A simple example (there are many other theories in this class, but  we will not discuss them in this work) of such theories is  the SU$(2)$ YM theory with five adjoint Weyl fermions, a theory belonging to a general class that we refer to as ``QCD(adj)".  
 In \S\ref{qcdadjoint}, we show  that the non-perturbative physics of this class of theories is amenable to a semiclassical treatment at any size of ${\bf S}^1$. Within the semiclassical approximation, which is believed to be arbitrarily accurate at weak coupling,  a bound on  the mass gap in the gauge sector is given by (with $N=2$ for this example): 
\begin{equation}
m (L)  <   \frac{c}{L} \; \; 
{\rm exp} \left[ {- \frac{8 \pi^2} {g^2_* N} } \right] ,  \qquad  {\rm for ~ any }  \;\; L   ~,
\label{mL}
\end{equation}
where $g_*$  is the fixed point of the renormalization group (RG) flow and $c$ is a  factor which may also depend on $g^2$ in a  power-law manner.
  The inverse of the mass gap (\ref{mL}) for gauge fluctuations 
  is the characteristic scale of gauge fluctuations  $\sim L  \;
{\rm exp} \left[ {\frac{8 \pi^2} {g^2_* N} } \right] $. 
  The fact that the inverse mass gap increases with increasing $L$ tells us that 
  the topological excitations  will eventually  dilute away to zero.  
   The mass gap goes to zero in the decompactification limit, as opposed to small-$N_f$ gauge theories for which topological excitations become non-dilute at $L \Lambda \sim 1$. For a theory where (\ref{mL}) holds   the   IR-CFT option should be realized according to our mass gap criterion (\ref{IRCCC}). 
      
\begin{figure}[t]
\centering
\includegraphics[width=4.5in]{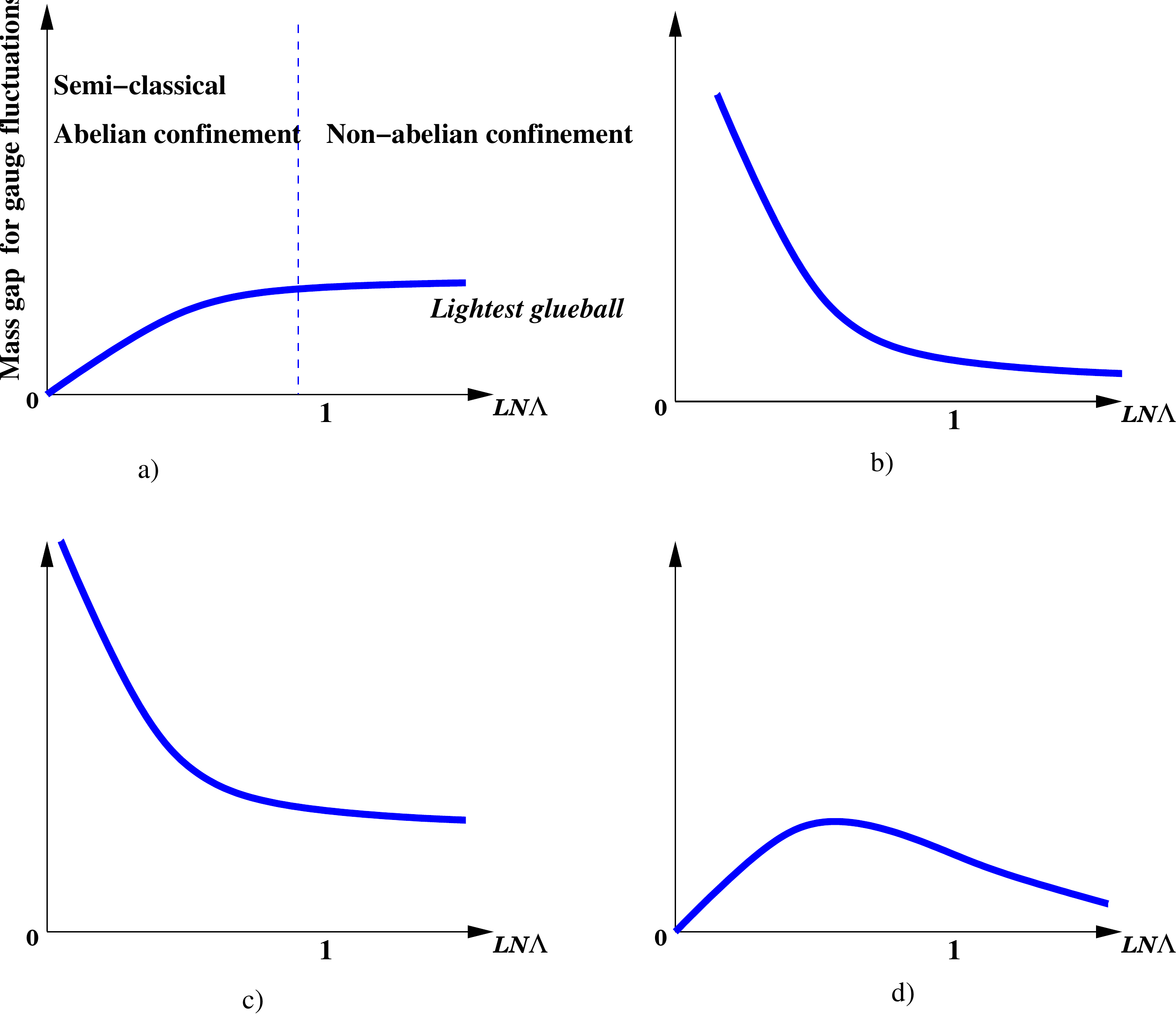}
\caption{Possible behavior of the mass gap for gauge fluctuations in asymptotically free, center-symmetric theories as a function of the radius of $\bf{S}^1$:  {\bf a.)}   $N_f$ small: mass gap increases in the semiclassical domain of abelian confinement and saturates to its ${\bf R}^4$  value in the non-abelian confinement domain. {\bf b.)} $N_f$  sufficiently large, perhaps  just below $N_f^{AF}$: mass gap is a decreasing function of radius. There are theories in this class for which semiclassical analysis applies at any size 
${\bf S}^1$. {\bf c.)}  Mass gap may start decreasing in the semiclassical domain, but may possibly saturate to a finite value  on  ${\bf R}^4$. This may happen, for example, if $\chi$SB takes place on the way.  {\bf d.)} Mass gap starts to increase in the semiclassical domain, but before reaching $\Lambda L \sim 1$, the coupling reaches a fixed point value without triggering $\chi$SB and the mass gap decreases to zero at larger scales. 
We will argue that {\bf a.)} and {\bf b.)}  occur at small  and large $N_f$, respectively, in all classes of theories we consider. We do not know whether {\bf c.)} and {\bf d.)} occur in any of the theories we consider and our semiclassical methods are of no help in this regard. 
We note, however, that  {\bf c.)} and {\bf d.)} are mutually exclusive---if either kind of behavior exists in a given class of theories for some $N_f$, the other kind is not expected to occur, see the text.
}
 \label{fig:massgap}
 \end{figure}

Motivated by this calculable example,  the  strategy behind our estimates of the conformal window is as follows.  For each theory, at small $L$ we  can  reliably estimate the mass gap for gauge fluctuations---the mass of the dual photon(s). Most important for our estimate of the conformal window boundary is 
 the behavior of the mass gap as a function of $L$ while keeping the strong-coupling scale of the theory $\Lambda$ the same for all $N_f$, in order to compare theories with different fermion content. 
 We will show that for small number  of fermion representations, $N_f$, the mass gap is always an increasing function of $L$. However, when   the number of fermions is increased beyond some value, $N_f^*$, the mass gap becomes a decreasing function of $L$. Our estimate of the lower bound of the conformal window is the value of $N_f^*$ when this change occurs. The intuition about the behavior of the mass gap and the possible caveats are illustrated on fig.~\ref{fig:massgap} and are further discussed below.

  \subsection{A conjecture: IR conformal or  confining on   ${\bf R}^4$?}
  \label{conjecture}
  
      As we saw above, our findings suggest  a way to understand why two gauge theories which are microscopically almost identical (for example, just above and  below  the conformal window) 
  may flow into drastically different infrared theories.  
We suggest that  in order to see the difference, we must either use the twisted partition function with a judicious choice of matter content so that  (approximate) center symmetry is preserved at any radius, or  stabilize the center symmetry externally by using center stabilizing double-trace deformations.  Once  this is done, 
  we expect that,  for IR-CFTs, the topological excitations become more   irrelevant with increasing the size of ${\bf S}^1$, and for confining theories, the topological excitations become more relevant with increasing size of ${\bf S}^1$, hence our title.    
     
    The behavior shown on fig.~\ref{fig:massgap}a.  is the one expected to occur in confining theories: the topological excitations become less dilute in the decompactification limit, causing confinement (and $\chi$SB, when it occurs). This is the behavior known to occur in all cases at sufficiently small $N_f$. Similarly, the mass gap behaves as in fig.~\ref{fig:massgap}b.  for sufficiently large $N_f$, for example close to the asymptotic freedom boundary, where the IR fixed point of the two-loop beta function is at  small coupling.
 
 We do not know whether the  behaviors of the mass gap shown on figs.~\ref{fig:massgap}c. or \ref{fig:massgap}d.  occur in any of the multi-flavor vectorlike or multi-generation chiral gauge  theories considered in this paper. The first case, fig.~\ref{fig:massgap}c.,  would imply that while the mass gap decreases with $L$ in the semiclassical regime, near  $\Lambda L \sim 1$ multifermion operators due to various kinds of monopoles  (and/or gluon exchange)  become sufficiently strong  to trigger dynamical symmetry breaking and change the behavior of the gauge coupling at large scales. The second case, \ref{fig:massgap}d., would imply that 
 while the mass gap increases with $L$ for $L \Lambda \ll 1$, as $L$ is increased  the gauge coupling reaches the fixed point value, without 
 causing any condensates to occur, and the large-$L$ scaling of the mass gap   in the decompactification limit is then as in eqn.~(\ref{mL}).
 
The  natural expectation is that in a given class of theories (same gauge group but different number of fermion flavors or generations, $N_f$) either \ref{fig:massgap}c. or \ref{fig:massgap}d. can occur for some $N_f$, but not both. This is because at small $L$ and $N_f$,  the  mass gap always increases with $L$ (as we  show in a controllable manner in later sections). Furthermore, since in $\bf{R}^4$ the small $N_f$ theories are known to confine, \ref{fig:massgap}a. is the behavior realized at small $N_f$. Now, upon increasing $N_f$, it may happen that \ref{fig:massgap}d. occurs before $N_f^*$ is reached (in such cases, our estimate $N_f^*$ for the conformal window boundary would be an upper bound thereof). Thus, a fixed point value of the coupling is reached without causing the dynamical breaking of any symmetry and the theory becomes conformal at some $N_f < N_f^*$. Upon increasing $N_f$ past $N_f^*$, one expects that   the mass gap  will eventually behave as on \ref{fig:massgap}b., as the theory which does not break the global symmetry and confine at smaller $N_f$ is not expected to do so as $N_f$ is increased, hence \ref{fig:massgap}c. is not expected to occur for any $N_f$ if \ref{fig:massgap}d. occurs. 

Conversely, at small $L$ the mass gap is always a decreasing function of the radius at $N_f> N_f^*$  and   at sufficiently large $N_f$ it is expected that \ref{fig:massgap}b. is always realized. Now, if \ref{fig:massgap}c. occurs, the theory confines upon decrease of  $N_f$ before it reaches $N_f^*$. But such a theory is expected to remain confining as $N_f$ is further decreased, i.e. \ref{fig:massgap}d. is not expected to occur if \ref{fig:massgap}c. occurs.  In such cases, our estimate $N_f^*$ for the conformal window boundary would be a lower bound thereof. 
 
 We note that the methods of this paper---as well as any theoretically controllable analytic methods---are not useful for deciding whether \ref{fig:massgap}c., \ref{fig:massgap}d. are actually  realized in any  of the   theories we consider in this paper.\footnote{For vectorlike theories, it would be possible to combine our analysis with truncated Schwinger-Dyson equations with monopole and/or gluon kernels at finite $L$ (however, all theoretical uncertainties mentioned in \S\ref{SDeqns}  apply here as well).} However,  near-future lattice studies will be able to shed light on this question, at least in the vectorlike case.

{\bf \flushleft{C}omparison with the case of thermal compactifications:} At asymptotically high temperature fermions decouple  and the long-distance theory is pure YM. The center symmetry is always broken, and the nonperturbative mass gap (``magnetic mass") is  of order $g_3^2 \sim T g_4^2(T)$. For sufficiently high $T$ (or small $L \sim T^{-1}$), the mass gap is always an increasing function of $T$, for confining and conformal theories alike. Note that this means that in thermal compactifications only the small-$L$ behavior of the mass gap of fig.~\ref{fig:massgap}b. and \ref{fig:massgap}c. occur, with the mass gap increasing with $L$, as opposed to the nonthermal compactification small-$L$  behavior we find. The thermal compactification is of no use in our diagnostic of conformality or confinement.

  \subsection{Estimates for the conformal window}
 \label{intro-estimates}
 
In this section, we present our results for $N_f^*$ for a sample of theories and a give a  brief comparison with the results of 
other approaches. These results, as well as those for other theories will be discussed in more detail in 
later sections.  

For all center-symmetric gauge theories of our interest on  small ${\bf R}^3 \times {\bf S}^1$, 
    the   mass gap for gauge fluctuations   is  of the form:
   \begin{equation}
m (L)  \sim     \frac{1}{L}
{\rm exp} \left[ {- q \frac{8 \pi^2} {g^2 (L)  N} } \right] ,
\label{mL2} 
\end{equation}
where the number $q \sim {\cal O}(1)$ depends on the theory (see  Table~\ref{default}, where $S_0 = 8 \pi^2/(g^2(L) N)$). 
For confining theories, we can convert the $e^{-8 \pi^2/( g^2(L)N)}$ factor to a strong scale  by using dimensional transmutation, via the one-loop beta function given in eqn.~(\ref{oneloop2}) (we do not demand the use of two-loop beta function due to the crude nature of our estimates).  Of course, 
  in an IR-CFT, there is really no dynamically generated  strong scale and the  scale introduced by the one-loop beta function   is  a fictitious  one.\footnote{\label{cftscale}Strictly speaking, there is also a scale in IR-CFT, which is the length scale $L_* \sim 1/\Lambda$ at which the running coupling is saturated to its infrared value. In our $L$-scaling of the mass gap, in order to compare different theories, we always keep $\Lambda$ fixed as $N_f$ changes.} By using the one-loop definition of the strong scale, we then express (\ref{mL2}) as:
  \begin{equation}
m (L)  \sim  \Lambda (\Lambda L)^{P(N,  N_f, {\cal R})}~. 
\label{mL3}
  \end{equation}
  If $P > 0$,  the mass gap increases with $L$ and 
  we suggest that this should lead to a confined theory, 
  and for  $P<0$ theories, we propose an IR-CFT. For confining theories, the region of validity of our semiclassical analysis is restricted to the $\Lambda L \ll 1$ domain. We then expect (based 
  on the wisdom gained from lattice gauge theory) that the mass gap should be saturated to a value  of order  $\Lambda$.  As already mentioned, for at least a subclass of IR-CFTs, the semiclassical analysis 
  is valid at any radius of the  ${\bf S}^1$ circle. 
  
  Below we present the results for a sample of vectorlike and chiral gauge theories. By 
  QCD(F), we denote an SU$(N)$ gauge theory with $N_f^D$ fundamental Dirac flavors,  QCD(S)---an SU($N$) theory with $N_f^D$ two-index  symmetric representation Dirac fermions,  and $[ {\rm S},  (N+4) {\overline F} ]$---an SU$(N)$ chiral gauge theory  with $N_f^W$ generations of two-index  symmetric tensor and $N+4$ antifunamental Weyl  fermions.
 Performing the calculation of the mass gap (\ref{mL2}) and of the exponent $P(N,  N_f, {\cal R})$ in (\ref{mL3}), our estimates of the conformal window  are:
  \begin{eqnarray}
  \frac{5}{2}N <  &N_f^D & < \frac{11}{2}N, \qquad \qquad \qquad  \;\;\;\;  {\rm QCD(F) } \cr\cr
   4 \left( 1 - \frac{2}{N+2} \right)  < &  N_f^D  & < \frac{11}{2}  \left( 1 - \frac{2}{N+2} \right) ,
   \qquad    {\rm QCD(S)}  \;\;    N \geq 3 \cr \cr
  4 \left( 1 - \frac{3}{N+3} \right)  <  & N_f^W & < \frac{11}{2}  \left( 1  -  \frac{3}{N+3} \right) , 
   \qquad   [  {\rm S},  (N+4) {\overline F} ] . \;\; N \geq 5 
   \label{resultsintro}
 \end{eqnarray}
 We   discuss 
 other interesting vectorlike and chiral theories in the rest of the paper.
 
 Interestingly, our results for the conformal window for  2-index representation vectorlike theories are almost coincident with the ladder approximation approach 
\cite{Appelquist:1996dq, Dietrich:2006cm} 
($| N_f^{*}({\rm ladder}) -  N_f^{*}({\rm n.p.}) | < 0.15,  \; 3 \leq N \leq \infty$),  but differ by some amount  from the SUSY-inspired  approach of  \cite{Ryttov:2007cx} (if the $\gamma = 2$ criterion is used).   For fundamental fermions, 
on the other hand, our non-perturbative estimate is much closer to the  NSVZ-inspired approach (which gives $ N_f^* = 2.75 N$  \cite{Ryttov:2007cx}), than the older estimates based on two-loop beta functions and the gap equation of \cite{Appelquist:1988yc, Appelquist:1996dq, Dietrich:2006cm} which yield a value slightly less than $ N_f^* \sim  4 N$. We can thus summarize our findings for vectorlike theories as follows:\footnote{Note, however, the work of \cite{Gies:2005as}, which uses the three- and four-loop beta function and the average action formalism to yield an estimate closer to (or perhaps slightly higher than)  $N_f^*\sim 3 N$.}
 \begin{eqnarray}
 {\rm Conformal \; \; window}\Big|_{\rm Deformation \; theory} &\approx &
  \left\{
\begin{array}{l l}
{\rm ladder \; approx.}\; (\gamma = 1) &\qquad   2{\rm -index \; reps.}  \cr 
{\rm susy-inspired} \; (\gamma \le 2) & \qquad {\rm fundamental \; rep.}
   \end{array} \right. \\
   &\ne &\left\{
\begin{array}{l l}
{\rm ladder \; approx.} \; (\gamma = 1) &\qquad    {\rm fundamental \; reps.}  \cr 
{\rm susy-inspired} \; (\gamma \le 2) & \qquad 2{\rm -index \; reps.} 
   \end{array} \right. \nonumber
 \end{eqnarray}
 
 This result is rather surprising.   As reviewed earlier, the ladder approximation approach relies 
 on a truncated Schwinger-Dyson equation for the 
fermion self energy,  a beta function usually at two-loop order, and 
 a large anomalous dimension of the fermion bilinear $(\gamma \sim 1)$ triggering $\chi$SB. The SUSY-inspired approach 
 uses an NSVZ-type beta function for QCD and the lower unitary bound on the dimension of scalar operators   $(\gamma \le 2)$. 
On the other hand, our main result relies on knowledge of the topological excitations which lead to confinement and mass gap in  the semiclassical regime on  ${\bf R}^3 \times {\bf S}^1$ and we use only the one-loop beta function. A priori, even  the fact that these approaches produce results in the same  ball-park is surprising, and we comment on some of
 the reasons in \S\ref{comparison}.

\subsection{Outline}
 
 The lines of thought leading to our conformal window bounds and some of our results were already described in the   Introduction above. The reader interested primarily in the    results should jump to \S\ref{comparison} where Tables~\ref{defaultChi}, \ref{defaultF}, \ref{defaultS}, \ref{defaultAS}, \ref{defaultADJ} present all our bounds and comparison with other estimates.
 
In the remainder of the paper, we give a more detailed discussion of the calculation of the mass gap in several classes of gauge theories, present the results for all other theories that we consider, and compare our findings to those of other approaches.  

We begin, in \S\ref{qcdadjoint}, by considering SU$(N)$ gauge theories with $N_f^W$ Weyl fermion adjoints  (this is one of the few classes of theories for which we consider the derivation of our bounds on the conformal window in some detail). We first review the special properties of adjoint theories.
In \S\ref{twisted}, we review the twisted partition function, the dynamical center-stabilization, and the main steps in the derivation of the long-distance effective theory for the   $N=2$ case, leading to the semiclassical estimate of the mass gap for the gauge fluctuations. In \S\ref{region}, we discuss the region of validity of the semiclassical analysis and argue that the $N_f^W=5$ theory is solvable semiclassically at any value of $L$. The mass gap for the adjoint theories is considered  in \S\ref{massgapadj} and the bounds on the conformal window are derived via its small-$L$ behavior. We also make some comments relevant to future lattice studies of  $N_f^W = 5$ adjoint theories. 

Next, we move to more general theories by first summarizing, in \S\ref{classification}, the results of previous work for the mass gap at small $L$, for a variety of theories. Table \ref{default} contains the main results used to obtain the conformal window estimates. In \S\ref{non-selfduality}, we emphasize the importance of the   novel composite topological excitations and their non-selfduality.

In \S\ref{vectorlike}, we consider vectorlike (``QCD-like") gauge theories. In \S\ref{su3symmetric}, we study in some detail the infrared theory of an SU$(3)$ gauge theory with one sextet-representation Dirac fermion, as it has not been previously considered in the literature (this theory is confining). In \S\ref{suNsymmetric}, we give our results for the conformal window for SU$(N)$ vectorlike theories with $N_f^D$ two-index symmetric Dirac fermions.
In \S\ref{suNfundamental} we study SU$(N)$ vectorlike theories with $N_f^D$ fundamental flavors. Section \ref{suNbifundamental} considers SU$(N)\times$SU$(N)$ theories with $N_f^D$ copies of bifundamental Dirac fermions and
\S\ref{suNantisymmetric}---SU$(N)$ theories with $N_f^D$ copies of two-index antisymmetric tensor Dirac fermions.

Chiral gauge theories are considered in \S\ref{chiral}. In \S\ref{chiralAS}, we consider SU$(N)$ theories with $N_f^W$ ``generations" of two-index antisymmetric tensor and $N-4$ antifundamental Weyl fermions, while similar theories with two-index symmetric tensors and $N+4$ fundamentals are considered in \S\ref{chiralS}. Finally, chiral SU$(N)^K$ quiver gauge theories are the topic of \S\ref{chiralquiver}. 

Comparison of our results for the conformal window with those of other analytic approaches are given in 
\S\ref{comparison}. We begin, in \S\ref{sdcomparison}, by giving comparisons of our approach with the truncated Schwinger-Dyson equations and supersymmetry-inspired estimates, see Tables~\ref{defaultF}, \ref{defaultS}, \ref{defaultAS}, \ref{defaultADJ}. In  \S\ref{universality}, we compare some features of the various approaches. 
In \S\ref{lattice}, we compare with  the available lattice data. 

In \S\ref{conclusions}, we conclude, discuss future directions and the possible relation of our work to recent work on the conformal-confining transition in multi-flavor QCD.
Appendix \ref{conventions}  summarizes some formulae for the beta functions and the strong scale that we use.

\section{Twisted partition function and QCD with $\mathbf{N_f^W}$ adjoint Weyl fermions}
\label{qcdadjoint}

As discussed in the   Introduction, Yang-Mills theories  with 
$N_f^W < 5.5$ massless adjoint Weyl  fermions, which we call QCD(adj),  may be useful in some  extensions of  the standard model, for example,  $N_f^W = 4$ may be a theory exhibiting  near conformal (``walking") or  conformal behavior.   This theory has few other unique properties which make it perhaps the most special and analytically tractable one among all vectorlike theories.   

\begin{itemize}
\item{ {\bf Unbroken (spatial) center symmetry:} With periodic boundary conditions for fermions, these theories never break their (spatial) 
center symmetry when formulated on ${\bf R}^{4-d} \times {\bf T}^d$, and in particular, on
${\bf R}^{3} \times {\bf S}^1$, and ${\bf T}^4$.} 
\item{ {\bf Large-N volume independence:} In the large-$N$ limit, the non-perturbative physics of QCD(adj) formulated on ${\bf R}^{4-d} \times {\bf T}^d$ is independent of the size of $d$-torus ${\bf T}^d$, i.e, it satisfies volume independence.}
\end{itemize}
The volume independence property at $N=\infty$ is an {\it exact} property of this class of gauge theories. It states, in particular, that if one can solve the  reduced  matrix quantum mechanics problem on arbitrarily small ${\bf R} \times {\bf T}^3$, it also implies the solution of the QCD(adj) on ${\bf R}^{4}$. In this paper, we will typically work with small $N$, and use exclusively the first property. 
The first property states that if one implements the chiral limit of this theory on the lattice and employs periodic boundary conditions for fermions, the analog of confinement-deconfinement transition of the thermal counterpart of this theory does not take place.

Studying the two-loop beta function for QCD(adj) for $N_f^W <5.5$, we see that there is a fixed point for $N_f^W= 3,4,5$ at, respectively,  couplings that can be characterized as  strong, intermediate, and weak.  The numerical values of the fixed-point couplings are: 
\begin{eqnarray}
\alpha^* = {g^2_{*}\over 4 \pi}=   - \frac{4 \pi  \beta_0}{\beta_1}  = \{ {\rm  none, \;  none}, \; 2.1, \; 0.6, \; 0.13\}, \qquad {\rm for} \: N_f^W= \{1,2,3,4, 5\} ~.
\label{fpalpha}
\end{eqnarray}
In this section, we will explore another magical property of QCD(adj) with $N_f^W=5$:
\begin{itemize}
\item{ {\bf Semi-classical solvability at any  size of ${\bf S}^1$:} Modulo a plausible assumption,  see the discussion in \S\ref{region},
the SU$(N)$  gauge theory with $N_f^W=5$ fermions  is  solvable at any size of ${\bf S}^1$ radius while remaining within a reliable semiclassical domain. }
\end{itemize}
This theory gives the first example of a non-supersymmetric non-abelian gauge theory for which the  
decompactification limit can be taken in a reliable fashion. We explain this in the following sections.

\subsection{Review of the  twisted partition function, one-loop potential, and magnetic bions} 
\label{twisted}

We first briefly review the  dynamics of the non-supersymmetric QCD(adj) formulated on    ${\bf S}^1 \times {\bf R}^3$ (with fermions endowed with periodic spin connection) in its semiclassical 
domain.  The aspect which makes this theory rather special is that adjoint fermions with periodic boundary conditions stabilize the center symmetry even at small   ${\bf S}^1$.  We are interested in the twisted partition function (\ref{twistedz}), which may be written as:
\begin{equation}
\label{twistedZ}
{\widetilde Z}(L)= \tr \left[ e^{-LH} (-1)^F \right] =   \sum_{n \in {\cal B}} 
e^{-LE_n}  - \sum_{n \in {\cal F}} 
e^{-LE_n}   ~,
\end{equation}
where ${\cal B}$ and ${\cal F}$ are the bosonic and fermionic Hilbert space of the gauge theory.   The microscopic theory possesses a global $( SU(N_f^W) \times \Z_{2N_f^W N})/ ( \Z_{N_f^W} \times  \Z_{2})$ chiral symmetry, where $\Z_{2N_f^W N}$ is the anomaly-free discrete subgroup of the classical axial $U(1)$ symmetry and $\Z_2$ is the fermion number symmetry (the denominator is the factored out symmetry to prevent double counting; the genuine discrete chiral symmetry of the theory is only the $\Z_{N}$ factor).  

For unbroken center symmetry, it is crucial that one employs a circle compactification with periodic boundary conditions.  Otherwise,  for thermal (anti-periodic) boundary conditions and  at sufficiently high temperatures, the center symmetry always breaks down and the theory moves to a  deconfined phase.  Fermions are gapped  due to thermal mass 
and the long distance theory theory reduces to pure Yang-Mills theory in 3 dimensions. Although interesting on its own right, due to the fact that this theory is not continuously connected to the 
gauge theory on  ${\bf R}^4$, it is not of particular interest to us in this paper. The difference 
in the center symmetry realization can be best understood by using 
the density matrix: in the thermal case, the density matrix is positive definite, and specifically at large $N$, the Hagedorn growth is indicative of some instability. With the use of the twisted 
partition function (\ref{twistedZ}), the ``twisted density matrix" is no longer positive definite. It is positive definite in the bosonic Hilbert space and negative definite in the fermionic one, 
$\rho(E)=    \sum_{n \in {\cal B}} 
\delta (E- E_n)  - \sum_{n \in {\cal F}} 
\delta(E-E_n)$.  This is how the circle compactification avoids the center symmetry change.  
 
 The stability of center symmetry even at arbitrarily small radius 
 is a zero-temperature  quantum mechanical effect, which can only occur in spatial circle compactification. It is not surprising that the physics of the small  ${\bf S}^1$ phase depends on boundary conditions. The  difference  as a function of boundary conditions  reflects the distinction between the thermal and quantum fluctuations in the  gauge theory.

Let $\Omega(x)= e^{i \int A_4(x,x_4) dx_4}$ denote the holonomy of the Wilson line along the compact spatial direction.  
 Considering, for simplicity, the  $SU(2)$ case, it can be brought into a diagonal form by a gauge rotation: 
\begin{equation}
\Omega(x)= \left( \begin{array}{cc}
 e^{i v} & 0 \\ 
 0 & e^{-iv} 
 \end{array} 
 \right)~.
 \end{equation}
In an appropriate range of  $L$,  where the gauge coupling is small, 
we may evaluate the one-loop effective potential for the spatial Wilson line reliably.   The result, see the appendix of \cite{Shifman:2008ja}, is: 
\begin{equation}
V^{+}[\Omega] = (-1+ N_f^W) \frac{2}{\pi^2 L^4} \sum_{n=1}^{\infty} \frac{1}{n^4}  |\tr \Omega^n|^2 
\label{pot1}
 \end{equation}
 where the $(-1)$ factor is due to  gauge fluctuations and the $(+N_f^W)$ term is the fermion-induced center-symmetry 
 stabilizing term.\footnote{As emphasized earlier, this stabilization is not possible with thermal boundary conditions. In that case, the one loop potential is:
 \begin{equation}
V^{-}[\Omega] =  \frac{2}{\pi^2 L^4} \sum_{n=1}^{\infty} \frac{1}{n^4}  (-1 + (-1)^n N_f^W) |\tr \Omega^n|^2 ~,
\label{pot2}
 \end{equation}
 and both the fermion-  and gauge boson-induced terms prefer the breaking of center symmetry.}
 There are $\O(g^2)$ corrections to this formula which are negligible so long as the coupling constant remains small.    The minimum of the one-loop  potential (\ref{pot1}) for $N_f^W > 1$ is located at $v= \pi/2$, thus center symmetry is intact at small ${\bf S}^1$, and plausibly at any value of   $L$: 
 \begin{equation}
\langle \Omega \rangle = \left( \begin{array}{cc}
 e^{i \pi/2} & 0 \\ 
 0 & e^{-i\pi/2} 
 \end{array} 
 \right) ~.
  \label{vev}
\end{equation}

Since the holonomy (\ref{vev}) behaves as an adjoint Higgs field,   the separated eigenvalues lead to 
gauge symmetry breaking $SU(2) \rightarrow U(1)$ at a scale $\sim L^{-1}$. At small $L$, the long-distance theory is that of a free photon. The 3d photon  is dual   to   a free 3d scalar field, the dual photon $\sigma$. Furthermore, there remain $N_f^W$ massless   fermions $\lambda^I$, neutral under the unbroken U$(1)$---the adjoint-fermion components which do not obtain mass due to the expectation value (\ref{vev}).
Thus,  the long-distance perturbative physics of the QCD(adj) theory at small $L$ is described by a  free field theory of $\sigma$ and $\lambda^I$, $I=1,...,N_f^W$.
 
 Nonperturbative effects, however, change this picture. It is by now well understood that  due to gauge symmetry breaking via the holonomy (a compact Higgs field), there  are two types of topological excitations, the BPS monopoles, which we denote by ${\cal M}_1$,  and KK monopoles, denoted by ${\cal M}_2$ \cite{Lee:1997vp,Kraan:1998pm}. It is also well-known that in pure YM theory, these can ``disorder" the system by generating a mass gap for the dual photon  by Debye screening in the monopole/anti-monopole plasma.  The mass gap of the dual photon causes  confinement of external electric charges and the string tension is of order the mass squared. As we already stated, the calculation of the mass gap is under theoretical control in  center-symmetric small-$L$ theories. 

However, in the theory at hand, due to the existence of zero modes of the adjoint fermions, the BPS and KK monopoles cannot induce a mass gap for gauge fluctuations.
Instead, they give rise to the following   monopole operators in the long-distance theory: 
   \begin{eqnarray}
 &&  {\cal M}_{\rm 1} = e^{-S_{0}} e^{i \sigma}  \det_{I, J} \lambda^I \lambda^J,  \qquad  \overline {\cal M}_{ 1} = 
   e^{-S_{0}} e^{- i \sigma}  \det_{I, J} \bar \lambda^I \bar \lambda^J ,  \cr \cr 
&&    {\cal M}_{\rm 2} = e^{-S_{0}} e^{- i \sigma}  \det_{I, J} \lambda^I \lambda^J , \qquad 
     \overline {\cal M}_{2 } =  e^{-S_{0}} e^{ i \sigma}   \det_{I, J} \bar \lambda^I \bar \lambda^J ,  \;
     \label{mon}  \qquad I, J=1, \ldots N_f^W~, 
   \end{eqnarray}
where $\overline{{\cal M}_1}$ and $\overline{{\cal M}_2}$ denote the operators generated by the anti-BPS and anti-KK monopoles, respectively,  $S_0$ is the monopole action given in  (\ref{instanton4}) below, and $\sigma$ is the dual photon field. 
The number of 
fermionic zero modes for these two topological excitations is dictated by the relevant index 
theorem   \cite{Nye:2000eg, Poppitz:2008hr} on ${\bf R}^3 \times {\bf S}^1$ and are given by:
\begin{equation}
{\cal I}_{\rm BPS} =2N_f^W, \qquad   {\cal I}_{\rm KK} = 2N_f^W, \qquad    {\cal I}_{\rm inst}= {\cal I}_{\rm BPS} + {\cal I}_{\rm KK} =2 N N_f^W= 4 N_f^W~. 
\end{equation}
    The 4d instanton operator may be viewed as a composite of these two types of monopole operators:  
   \begin{eqnarray}
   \label{instanton4}
 I_{\rm inst.} ={\cal M}_{\rm 1}   {\cal M}_{\rm 2}  =   e^{-2S_0} (\det_{I, J} \lambda^I \lambda^J)  (\det_{I, J} \lambda^I \lambda^J), 
   \qquad 
 S_{\rm inst} =     2 S_{0} =  \frac{8 \pi^2}{g^2}~,
    \end{eqnarray}
and is also unimportant for confinement at small $L$, as it is $\sigma$-independent.  

Now, note that under the anomaly-free chiral symmetry $\Z_{2N_f^W N}$, 
$\det_{I, J} \bar \lambda^I \bar \lambda^J \rightarrow - \det_{I, J} \bar \lambda^I \bar \lambda^J $, 
which is a  $\Z_2$ action.  
Thus, the invariance of the monopole operator demands that the dual photon  transform by a discrete shift symmetry:
\begin{equation}
  [\Z_2]_{*}: \qquad \sigma  \rightarrow \sigma + \pi~.  
  \label{z2star}
  \end{equation}
   The $[\Z_2]_{*}$ symmetry permits 
 purely bosonic  flux operators of the form:
   \begin{eqnarray}
 &&  {\cal B}_{\rm 1} = e^{-2S_{0}} e^{2i \sigma},  
  \qquad  \overline {\cal B}_{ 1} = 
   e^{-2S_{0}} e^{- 2i \sigma}  ~, 
        \label{bion}
   \end{eqnarray}
which are referred to as ``magnetic bions".  They can be viewed as due to composites of the elementary topological excitations, 
 ${\cal M}_{\rm 1}$ and $\overline   {\cal M}_{\rm 2}$. Despite the fact that these two excitations interact repulsively via the Coulomb law, the fermion zero mode exchange generates a logarithmic attraction, which leads to the stability of magnetic bions \cite{Unsal:2007jx}.

Summarizing the above findings, the theory dual to the $SU(2)$ QCD(adj) in the semiclassical regime can be written as:
\begin{eqnarray}
L^{\rm dQCD(adj)} = \frac{g^2(L)}{2 L} (\partial \sigma)^2 -  {b\over L^3}\;  e^{-2S_0}\cos 2 \sigma  
+  i \bar \lambda^I \gamma_{\mu} \partial_{\mu} \lambda_I   
 + {c\over L^{3 - 2 N_f}} \; e^{-S_0}  \cos \sigma   
( \det_{I, J} \lambda^I \lambda^J +  \rm c.c.) ~, \qquad \label{Eq:dQCD}
\end{eqnarray}
where $I, J=1, \ldots N_f^W$ are summed over; the numerical coefficients $b$ and $c$   can    contain power-law dependence on the gauge coupling $g^2(L)$, which is inessential for our estimates.  
The mass (squared) gap due to magnetic bions appears at second order $(e^{-S_{0}})^2$ in the semiclassical expansion.

\subsection{Region of validity and theories solvable   at any  size ${\bf S}^1 \times {\bf R}^3$}
\label{region}

    The range of validity of the one-loop potential (\ref{pot1}), leading to center-symmetry preservation,
 depends on the particulars of a  theory. Confining gauge theories on ${\bf R}^4$ possess a strong scale $\Lambda$. For such theories,  
 the  one loop analysis is reliable if the running coupling is small, i.e,  $L \Lambda \ll 1$. The  QCD(adj) theories with small number of flavors $N_f^W=1,2,3$ are of this type.  Plausibly, the   $N_f^W=4$ theory is also just below the conformal window or perhaps conformal.  

However, the  $N_f^W=5$ theory seems to have an infrared  fixed point  at weak-coupling.  In general, 
  for asymptotically free theories with a weak coupling infrared fixed point, 
 the region of  validity of (\ref{pot1}) and dual theory (\ref{Eq:dQCD}) extend  to all values of $S^1$ radius. Let    $g^2_*$  denote the  weak coupling fixed  point, reached at the length scale 
 $L_*$.    
 Since:
 \begin{equation}
g^2 (L)  <  {g^2_*} \equiv  g^2(L_*),  \qquad   {\rm  for \; all \; } L  ,
\label{coupling}
\end{equation}
and the loop factor is small, $\frac{g^2(L_*)}{4 \pi}= 0.13 \ll1 $, 
it seems plausible that the region of validity 
of the semiclassical analysis on  ${\bf R}^3 \times {\bf S}^1$ can be extended to arbitrarily large ${\bf S}^1$. Thus, the dual formulation  (\ref{Eq:dQCD}) of the  $N_f^W=5$ theory is valid at  any
$ 0 < L < \infty $ (of course, the dual theory only holds at energy scales below $1/L$). 

There is one caveat to this argument, which was also referred to in the Introduction.  The fixed-point value of the coupling constant in  YM theories with matter in two-index representations cannot be tuned to arbitrarily small values, unlike the Banks-Zaks limit where the coupling constant can be tuned in such a way that three- and higher-loop corrections to the beta function  are negligible.   The fixed point in QCD(adj) with  $N_f^W=5$  is at a small {\it finite} value of the coupling constant.  Thus, 
once the one-loop and two-loop beta function are balanced, the higher loop effects can also give sizable effects, and in this sense, there is no controllable expansion. 

On the other hand, there are examples in which the Banks-Zaks limit extends to a regime 
$\frac{N_f^*}{N} < \frac{N_f}{N} < \frac{N_f^{AF}}{N}$, where $\frac{N_f^*}{N}$  is at a finite distance from $\frac{N_f^{AF}}{N}$. For example, in SUSY QCD, this window is $\frac{3}{2} < \frac{N_f}{N} < 3$.   For some of these theories, the fixed point coupling is small and finite, and higher order effects 
in the beta function are not totally negligible. However, there is reason to believe that the higher effects do not destabilize the fixed point. In view of the small value of the coupling constant, we envision that this may be the case in  $N_f^W=5$ QCD(adj). Therefore, our assertion regarding the validity of our semiclassical solution for this theory should  be viewed as conjectural.  Nonetheless,  this conjecture may be tested on the lattice if sufficiently light fermions can be used in  simulations.

\subsection{Mass gap,  non-perturbative bounds, and conformal window}
\label{massgapadj}

As explained in \S\ref{intro-estimates}, 
using the one-loop beta function, $\beta_0$, we can rewrite the magnetic bion induced mass gap in terms of the strong coupling scale:
\begin{equation}
\label{betaqcdadjoint}
\Lambda^{b_0} = 
\left({1 \over L}\right)^{b_0} e^{- \frac{8 
\pi^2}{g^2{(L)}N}}, \qquad  b_0 \equiv \frac{\beta_0}{N} = \frac{11}{3} - \frac{2 N_f^W} {3},  \qquad {\rm or,}~ {\rm equivalently} \qquad 
e^{-S_0(L)}= (\Lambda L)^{b_0}~,
\end{equation}
 where, as in (\ref{coupling}), we denote by $g(L)$ the running coupling at the energy scale $1/L$.
Note that the scale $\Lambda$ may or may not be dynamically generated in the  gauge theory; for a CFT, $\Lambda \sim L_*^{-1}$, the scale where the coupling reaches the fixed-point value. 

 Next, using (\ref{betaqcdadjoint}), the bion-induced mass gap for the dual photon from the second term in  (\ref{Eq:dQCD}) can be cast in the form:
\begin{equation}
m_{\sigma} \sim \frac{1}{L} e^{-S_0(L)}=   \frac{1}{L} e^{-\frac{8 \pi^2}{g^2(L) N}} =\frac{1}{L} (\Lambda L)^{b_0} =  \Lambda(\Lambda L)^{(8-2N_f^W)/3} ~.
\label{Eq:Spec}
\end{equation} 
This expression is valid for all QCD(adj) theories in their semiclassical domain $\Lambda L \ll 1$ where abelian 
confinement holds.   For theories with a low number of flavors, the theory moves to a non-abelian confinement domain when  $\Lambda L \sim 1$  and one loses  analytical control 
over the semiclassical approximation.    However, one expects a mass gap for gauge fluctuations of the order of strong scale to saturate to a value of order the strong scale   $ \Lambda$. For the theories in the conformal window, as the $N_f^W=5$ QCD(adj),  the mass gap is a decreasing function of radius;  according to our conjecture, this theory will flow to an interacting IR-CFT  on ${\bf R}^4$.  

For the QCD(adj) class of theories, the mass gap at large ${\bf S}^1$  is expected to be:
 \begin{eqnarray}
&& m_{\rm gauge \; fluct.} (L) \sim
  \left\{
\begin{array}{lll}
  \Lambda & \;\;\;  L \gg \Lambda^{-1}, \;\;  N_f < N_f^{*}  & \qquad {\rm confined}   \cr 
 \frac{1}{L}
 {\rm exp} \left[ {- \frac{8 \pi^2} {g^2_* N} } \right] , 
   & \;\;\;    L \gg L_{*}, \;\;\;
       N_f^{*} <  N_f <  N_f^{AF}   &\qquad  {\rm  IR-CFT}
   \end{array} \right.
   \label{largeS}
 \end{eqnarray}
From eqn.~(\ref{Eq:Spec}), we observe that the change of behavior of the mass gap as function of $L$ occurs at $N_f^* = 4$, hence 
 the estimate for the conformal window in QCD(adj) that we obtain from our conjecture is: 
  \begin{equation}
  4  <  N_f^W  < \frac{11}{2}, \qquad  {\rm QCD(adj)}~.
 \label{cwa}
  \end{equation}

It is interesting to note that an IR-conformal field theory on ${\bf R}^4$ actually exhibits confinement  without $\chi$SB  when compactified   on  ${\bf R}^3 \times {\bf S}^1$ with small $L$.  However, 
the scale at which confinement  sets in is most likely invisible in lattice gauge theory, as the estimates given below show.  We think that if this theory is simulated on the lattice, instead of the confinement phase, an abelian Coulomb phase with massless neutral fermions and photons 
will be observed for a practical range of lattice parameters (provided the $W$-bosons of mass $\sim L^{-1}$  can be distinguished from the massless photons).  To see this, 
consider an asymmetric lattice which mimics the  ${\bf R}^3 \times {\bf S}^1$ geometry: 
\begin{equation}
{\bf T}^3 \times {\bf S}^1  \approx 
{\bf R}^3 \times {\bf S}^1  \qquad {\rm provided}  \qquad  r({\bf T}^3)  \gg L({\bf S}^1)  ~.
\label{lattice1}
\end{equation}
The correlation length of gauge fluctuations (\ref{Eq:Spec})  is: 
\begin{equation}
m (L)^{-1} \sim  L   \; 
{\rm exp} \left[ {+ \frac{8 \pi^2} {g^2_* N} } \right]  , \qquad  L \in (0, \infty)~. 
\label{mL5}
\end{equation}
As we discussed in the previous section, we expect that in the $N_f^W=5$ theory this result is valid at any $L$, including $L> L_{*}$. In the decompactification limit,  (\ref{largeS}) states that the mass gap for gauge fluctuations vanishes or the correlation length is infinite.  The fixed point of the beta function, see (\ref{fpalpha}), is approximately located at 
 $g_*^2 = 1.7$.
This means that confinement in this theory (formulated as indicated in (\ref{lattice1})) will set in at distances 
 $e^{\frac{8 \pi^2}{2 g_*^2} } L  \sim e^{23} L $. Although such a theory on 
 ${\bf R}^3 \times {\bf S}^1$ is in principle confining, in a  practical simulation performed on 
 ${\bf T}^3 \times {\bf S}^1$,  it is not possible to see this mass gap. Current lattice simulations are performed on lattices for which $ \frac{r({\bf T}^3)}{  L({\bf S}^1)} \sim {\cal O}(1-10) $. 
 Thus, a lattice gauge theorist simulating this theory must see an abelian Coulomb phase with massless photons,   massless fermions, and $W$-bosons of mass $~L^{-1}$.
  This means that  the topological $[\Z_{2}]_{*}$ symmetry (\ref{z2star}) enhances to an  emergent continuous $U(1)_J$ shift symmetry for the dual photon.

  On $\bf{R}^4$, we do not expect dynamical abelianization to take place at any length scale in 
  QCD(adj) with $N_f^W=5$. Rather, we expect the $W$-boson  components of the gauge fluctuations to remain massless as well. 
  We conclude that  all the topological excitations in this gauge theory, magnetic monopoles,  
  magnetic bions, and instantons  are  irrelevant in the renormalization group sense. 
  The long distance theory on $\R^4$ is described in terms of the short distance quarks and gluons, and the long distance lagrangian is same as classical lagrangian. The theory is in a non-abelian 
  Coulomb phase of interacting quarks and gluons. 
  To the best of our knowledge, this is the first non-supersymmetric non-abelian gauge theory example where non-perturbative dynamics can 
  be understood semiclassically  at any $L$, including the   
  decompactification limit.

\section{Classifying confinement mechanisms in vectorlike and chiral theories on 
$\mathbf{R^3 \times S^1}$}
\label{classification}

The essence of the recent  progress in asymptotically free  YM theories with or without  fermionic matter 
is that new quantitative methods to study the non-perturbative dynamics  were found. 
In particular, at small ${\bf S}^1$, and sometimes at any size  ${\bf S}^1$, 
it is possible to understand  the IR dynamics  by using semiclassical methods and the relevant 
index   theorem on  ${\bf S}^1 \times {\bf R}^3$ \cite{Nye:2000eg, Poppitz:2008hr}.     The  confinement mechanisms in various vectorlike and chiral theories can thus be understood in a controllable manner. For detailed discussions of these mechanisms,  see  \cite{Unsal:2007jx, Shifman:2008ja, Unsal:2008ch, Shifman:2008cx, Shifman:2009tp, Poppitz:2009kz}.

The above studies showed that the mechanism of confinement in center-symmetric gauge theories on  ${\bf R}^3 \times {\bf S}^1$ 
depends very much on the details of the theory. In Table \ref{default},  we provide a list of 
 confinement mechanisms in many interesting SU$(N)$ gauge theories with vectorlike and chiral matter; the results are obtained similar to the analysis for the SU$(2)$ QCD(adj) theory given in \S\ref{qcdadjoint}.  The notation  we use in Table \ref{default} and elsewhere in the paper to refer to the various classes of theories we consider is as follows:
\begin{itemize}
\item ``YM" denotes pure SU$(N)$ YM theory.
\item ``QCD(F)" is the vectorlike SU$(N)$ theory of $N_f^D$ Dirac fundamentals.
\item ``SYM/QCD(adj)" is the vectorlike SU$(N)$ theory of $N_f^W$ copies of  adjoint representation Weyl fermions, whose dynamics for $N=2$ is considered in detail in \S\ref{qcdadjoint}; SYM indicates that the $N_f^W=1$ theory has   4d ${\cal N}=1$ supersymmetry.
\item ``QCD(BF)" is a vectorlike SU$(N)\times$SU$(N)$ gauge theory with $N_f^D$ Dirac fermions in the bifundamental representation.
\item ``QCD(AS)" is the vectorlike SU$(N)$ theory of  $N_f^D$ two-index antisymmetric tensor Dirac fermions. 
\item ``QCD(S)" is the vectorlike   SU$(N)$  theory of $N_f^D$  two-index symmetric tensors Dirac fermions.
\item ``SU(2) YM $I = 3/2$" denotes the chiral three-index symmetric tensor Weyl-fermion theory \cite{Poppitz:2009kz}.   
\item ``chiral $[SU(N)]^K$" denotes an SU$(N)^K$ quiver chiral gauge theory with $N_f^W$ copies of Weyl fermion   bifundamentals under any two neigboring gauge groups, see \S\ref{chiralquiver}. 
\item ``$AS+(N-4)\overline{F}$" is the chiral  SU$(N)$ theory of $N_f^W$ copies of 
the  two-index antisymmetric tensor  and $N-4$ antifundamental representation Weyl fermions
\item ``$S+(N+4)\overline{F}$" denotes the chiral  SU$(N)$ theory with $N_f^W$ copies of the  two-index symmetric tensor  and $N+4$ antifundamental representation Weyl fermions
\end{itemize}
  
  \begin{table}[h]
\begin{center}
\begin{tabular}{| p{2.7cm} |p{2cm}|p{3.5cm}|p{3.5cm}|c|}
\hline
  Theory  & Confinement mechanism on $\R^3 \times S^1$ & Index for monopoles 
 $ [{\cal I}_1, {\cal I}_2, 
   \ldots, {\cal I}_N ] $ 
  &  Index for instanton  \qquad $I_{inst.} = \sum_{i=1}^{N}  I_{i} \; \; \; \;  $
   &   (Mass Gap$)^2$   \\ \hline 
 YM  & monopoles   &  $[0, \ldots, 0]$  &  0&  $e^{-S_0}$  \\ \hline 
 QCD(F) & monopoles   & $ [2, 0, \ldots, 0] $  &  $2$ & $e^{-S_0}$  \\ \hline
 SYM/QCD(Adj) & magnetic bions  & $[2, 2, \ldots, 2]$   &  $2N$ & $e^{-2S_0}$  \\ \hline
  QCD(BF) & magnetic bions  &  $[2, 2, \ldots, 2]$   &$2N$ & $e^{-2S_0}$  \\ \hline
  QCD(AS) &   bions and  monopoles  &   $[2, 2, \ldots,2, 0, 0 ]$ 
   & $ 2N-4$ & $   e^{-2S_0}, e^{-S_0} $  \\ \hline
    QCD(S) &  bions and   triplets  &     $ [2, 2, \ldots,2, 4, 4 ]$ & 
    $ 2N+4 $ &
    $    e^{-2S_0}, 
    e^{-3S_0}  $  \\ \hline  \hline
      $SU(2) \;  {\rm YM} \;   I={3 \over 2}$ & magnetic quintets &     $[4,6]$ &  10& $ e^{-5S_0}   $  
      \\ \hline 
         chiral  $[SU(N)]^K$  & magnetic bions&     $[2,2,, \ldots,2 ]$ & $2N$ & $ e^{-2S_0}   $  
      \\ \hline 
            $AS + (N-4) {\rm \overline F}  $  & bions and a monopole &    
             $[1,1,, \ldots,1, 0,0 ] +  [0,0, \ldots,0, N-4,0 ]
            $ & $(N-2){\rm AS} + (N-4)  {\overline {\rm F}}  $ & $ e^{-2S_0}, e^{-S_0},    $  
      \\ \hline 
     $S + (N+4) {\rm \overline F}  $  & bions and triplets &    
             $[1,1,, \ldots,1, 2, 2] +  [0,0, \ldots,0, N+4,0 ] 
            $ & $(N+2){\rm S} + (N+4)  {\overline {\rm F}}  $ & $ e^{-2S_0}, e^{-3S_0},    $  
      \\ \hline 
\end{tabular}
\end{center}
\caption{
Topological excitations which determine  the mass gap for gauge fluctuations and  chiral symmetry realization 
   in vectorlike and chiral gauge theories on ${\bf R}^3 \times {\bf S}^1$. Unless indicated otherwise, all theories have an SU$(N)$ gauge group and their matter content was described earlier in this section. Confinement is induced by topological flux operators which do not have any fermionic zero modes. Monopole operators (with fermion zero mode insertions) cannot lead to confinement. In cases where most monopole operators carry fermion zero modes, the magnetic bions  generate the mass gap. However, monopole operators have typically fewer fermionic zero modes than the 4d-instanton, hence are more relevant for the chiral symmetry realization. The index theorems are given for $N_f=1$ ``flavors", for $N_f>1$, multiply the above results by  $N_f$.
}
\label{default}
\end{table}%

The knowledge displayed in Table \ref{default} for the various confinement mechanisms for center-symmetric theories on  ${\bf R}^3 \times {\bf S}^1$ at small $L$
 is at the heart of the main results of this paper and our  answer to the main question that we pose: on 
 ${\bf R}^4$, why does an  IR-confining gauge theory does confine and why does an IR-conformal theory flow to a CFT? 
  Crucial for this purpose is the order in the semiclasical expansion at which topological excitations give rise to the dual photon mass.    Using the one-loop result  $e^{-S_0(L)} \equiv (\Lambda L)^{b_0}$ ($b_0 = \beta_0/N$, see (\ref{betaqcdadjoint})),
we can determine the dependence of the mass gap on $N_f$ and $L$, at fixed $\Lambda$, in any of the gauge theories given 
above and apply our diagnostic.  The fact that for fundamentals, QCD(F), the mass gap is of order 
$e^{-S_0/2}$, while it is generically of order $e^{-S_0}$ in  two-index theories makes for important differences between our estimation and estimates based on the ladder approximation and the
NSVZ-inspired approach (when the $\gamma=2$ criterion is used).  This difference is of a non-perturbative nature (regarding the confinement mechanism) about which all-order summations of rainbow graphs  cannot account for. 

\subsection{Non-selfduality and  magnetic bions, triplets, quintets}
\label{non-selfduality}
As summarized   in Table~\ref{default},  a large class of novel topological excitations were discovered in non-abelian gauge theories  on ${\bf R}^3 \times {\bf S}^1$ during the last two years. These  excitations are referred to as magnetic bions \cite{Unsal:2007jx}, triplets (see \S\ref{su3symmetric} and figure \ref{fig:monop}), and quintets \cite{Poppitz:2009kz}. 
Perhaps, the most striking property of these excitations is that they are non-selfdual, unlike the monopoles or instantons.  However, they can be viewed as composite topological excitations, which are bound states of the selfdual and anti-selfdual excitations.
  Here, for completeness, we will say few words on this new class of topological excitations.  

The usual 't Hooft-Polyakov monopole and BPST instanton are solutions to the classical equations of motion. On ${\bf R}^3 \times {\bf S}^1$, the monopole-instantons are the solutions to the Prasad-Sommerfield (PS) equations, $\partial_k A_4 =  B_k$ with a non-trivial
holonomy,  $\Omega(x)= e^{i \int A_4(x,x_4) dx_4}$ at the boundary. This equation is the dimensional reduction of the  selfduality equation $F_{\mu \nu}= \half \epsilon_{\mu \nu \rho \sigma}  F^{\rho \sigma}$, whose solutions on   ${\bf R}^4$ are instantons.  

The magnetic bions, triplets, and other topological composites are {\it  not} solutions to the 
PS-type equations. However, they are permitted by symmetries and  are stable quantum mechanically. The common thread  of all these excitations is that they carry a net magnetic charge, but no fermionic zero modes. Hence, unlike the generic case with monopoles, they are 
able to produce confinement, in a regime where semiclassical techniques apply. 

Interestingly, the electromagnetic interaction between the  constituents of these composite excitations is {\it always} repulsive and one may not a priori expect them to form bound states. However, these objects have fermionic zero modes (the relevant indices are given in Table~\ref{default}). The zero modes induce attraction which overcomes the Coulomb repulsion between the constituents \cite{Unsal:2007jx}.\footnote{This effect may effectively be thought as coming from the Dirac operator. For example, in vectorlike theories,  if the   fermions are integrated out,    a term proportional to   ${\rm log \;  det} [\Dslash (A_\mu))]$ is induced in the action.  This term, in essence, provides the root cause for the stability of magnetic bions and other similar topological excitations.}

 The main lesson that we learned in the last few years is that non-perturbative aspects of both chiral and vectorlike theories are amenable to semiclassical treatment by using either twisted partitions function or deformation theory. In all cases, the cause of confinement is due to topological objects  which are non-(anti)selfdual.   The fact that these objects are non-selfdual is the main reason that they were not discovered earlier.\footnote{An important ingredient of the bions, triplets, and quintets---the twisted (or Kaluza-Klein) monopoles---was discovered only in 1997 \cite{Lee:1997vp,Kraan:1998pm}.}
 
 The selfdual objects, like monopoles, also play  a role in the dynamics, e.g. in the spontaneous breaking of global chiral symmetries.  At this end, there is also a crucial difference between chiral 
 and vectorlike theories. In all chiral theories but $SU(2)$ with $I= \frac{3}{2}$,   the monopole operators completely drop out of the dynamics due to averaging over global symmetries \cite{Shifman:2008cx}. We do not know if  this phenomenon may be tied with the complex phase of the fermion determinant in general chiral gauge theories.

 \section{Deformation theory and conformal windows in QCD with  complex representation}
\label{vectorlike}
 
 Our next  goal, similar to \S\ref{qcdadjoint}, is to address the dynamics of   asymptotically free Yang-Mills theories with $N_f^D <N_f^{AF}  $ Dirac  fermions in a complex  representations   of the gauge group, such as  QCD(F), QCD(BF), QCD(AS), and QCD(S) (see \S\ref{classification} for the definition of these theories).  Since these are all vectorlike theories with Dirac fermions in various representations $\cal{R}$, we refer to them collectively as
 ``QCD(${\cal{R}}$)" in what follows.
 
 We wish to understand the IR aspects of these theories, whether they yield   confinement or  conformality, and the non-perturbative aspects which lead to one or the other option.  As there are no current tools to address these questions directly on ${\bf R}^4$, we  use circle compactification to 
  ${\bf R}^3 \times {\bf S}^1$.  
 Once compactified on   ${\bf R}^3 \times {\bf S}^1$, none of the YM theories with complex representation fermions preserve  the  (approximate) center  symmetry at sufficiently small circle, regardless of whether one uses  periodic or anti-periodic boundary conditions. 
  This is unlike adjoint fermions. The reason that periodic boundary conditions for $ {\cal R} = \{
 \rm F, S, AS, BF\}$ cannot stabilize center symmetry is the misalignment of the color representations 
 of fermions and gluons. 
 However,  it was recently shown that it is possible to deform the QCD(${\cal R} $) in the small-circle regime so that  the deformed theory  smoothly connects to the large   ${\bf S}^1$ limit of the original theory, at least for   theories without continuous flavor symmetries.
  For gauge theories with massless multiple flavors, this deformation guarantees that there is no distinction between small- and  
    large-${\bf S}^1$ physics in the sense of center symmetry. However, in theories with continuous chiral symmetries, there may still be  $\chi$SB transitions on the way \cite{Shifman:2009tp}. 

The utility of  deformation theory is that it allows a controlled semiclassical analysis of the non-perturbative aspects of the theory, as was the case in QCD(adj) of \S\ref{qcdadjoint}.  We refer to the deformed QCD theories as QCD*.   
We assert that QCD$({\cal R})^{*}$ formulated on  ${\bf R}^3 \times {\bf S}^1$ can be used to deduce  aspects of  the  infrared behavior of QCD$({\cal R})$ on  ${\bf R}^4$. Following the same strategies as 
in QCD(adj), we will give an estimate of the conformal window via our mass gap criterion.  

As already noted, the confinement mechanisms of QCD(F/AS/BF)$^*$ for small number of flavors are already described in the literature \cite{Shifman:2008ja} and are given in Table~\ref{default}. QCD(S)* was  not studied in detail in ref.~\cite{Shifman:2008ja} due to its similarity to QCD(AS)* at large $N$. At small $N$, it exhibits a novel mechanism of confinement, which we discuss below. Hence, we begin by  first discussing the dynamics of QCD(S) with  $N_f^D  <   N_f^{AF}$  flavors. 

 \subsection{$\mathbf{SU(3)}$ QCD  with $\mathbf{N_f^D=1}$ sextet fermions}
 \label{su3symmetric}
 
SU$(3)$ QCD with two-index symmetric representation fermion has attracted the attention of some lattice studies recently, see \S\ref{lattice}.  Below, we will study this theory by using deformation theory. We start with $N_f^D=1$. Our discussion will be concise.

This theory has a $U(1)_V \times U(1)_A$ classical symmetry. The instanton has $(2N+4)N_f^D = 10$ zero modes, and hence, the axial symmetry of the quantum theory is $\Z_{10}$. 
In the weak-coupling small-${\bf S}^1$ regime of the center stabilized QCD* theory, the gauge symmetry $SU(3)$ reduces to $U(1)^2$ due to Higgsing by the adjoint holonomy. As before, since 
the holonomy is compact, there are 3 types of topological excitations, 2 BPS and 1 KK monopole, which we call ${\cal M}_{\rm 1,2,3}$, respectively. The index    theorem on ${\bf R}^3 \times {\bf S}^1$  \cite{Poppitz:2008hr} yields:
\begin{equation}
[{\cal I}_{1},   {\cal I}_{2},   {\cal I}_{3} ] = [4,4,2], \qquad       {\cal I}_{\rm inst}= \sum_{i=1}^3 {\cal I}_{i}~,
\end{equation}
zero modes per each monopole.  The three    monopole-induced operators are  (schematically):
   \begin{eqnarray}
  {\cal M}_{\rm 1}(x) = e^{-S_{0 }} e^{i \alpha_1\cdot \sigma} \psi^4, \qquad 
    {\cal M}_{\rm 2} = e^{-S_{0 }} e^{  i  \alpha_2 \cdot\sigma} \psi^{4}, \qquad 
   {\cal M}_{\rm 3} = e^{-S_{0 }} e^{  i \alpha_3 \cdot\sigma} \psi^{2}~. 
     \label{mon2}
   \end{eqnarray}
Here, $\alpha_1, \alpha_2$ are simple roots of the Lie algebra of $SU(3)$, $\alpha_3= - \alpha_1 -\alpha_2$ is the affine root, and $\sigma = (\sigma_1, \sigma_2)$ are the two dual photons. An explicit basis for the  affine root system is:
\begin{equation}
 \alpha_1= \left( \textstyle \frac{1}{2},  \sqrt{3\over 2}\right),  \qquad \alpha_2=\left (  \textstyle\frac{1}{2}, -\sqrt{3\over2}\right), \qquad 
 \alpha_3= ( \textstyle -1, 0)~.
 \end{equation}
 The charges of the magnetic monopoles under the unbroken 
$U(1)^2$ gauge group  are $\frac{4 \pi}{g} \alpha_i$. Similar to (\ref{mon}), the antimonopole operators are conjugate to those shown in (\ref{mon2}).

 \begin{figure}[t]
 \begin{center}
\includegraphics[angle=0, width=4in]{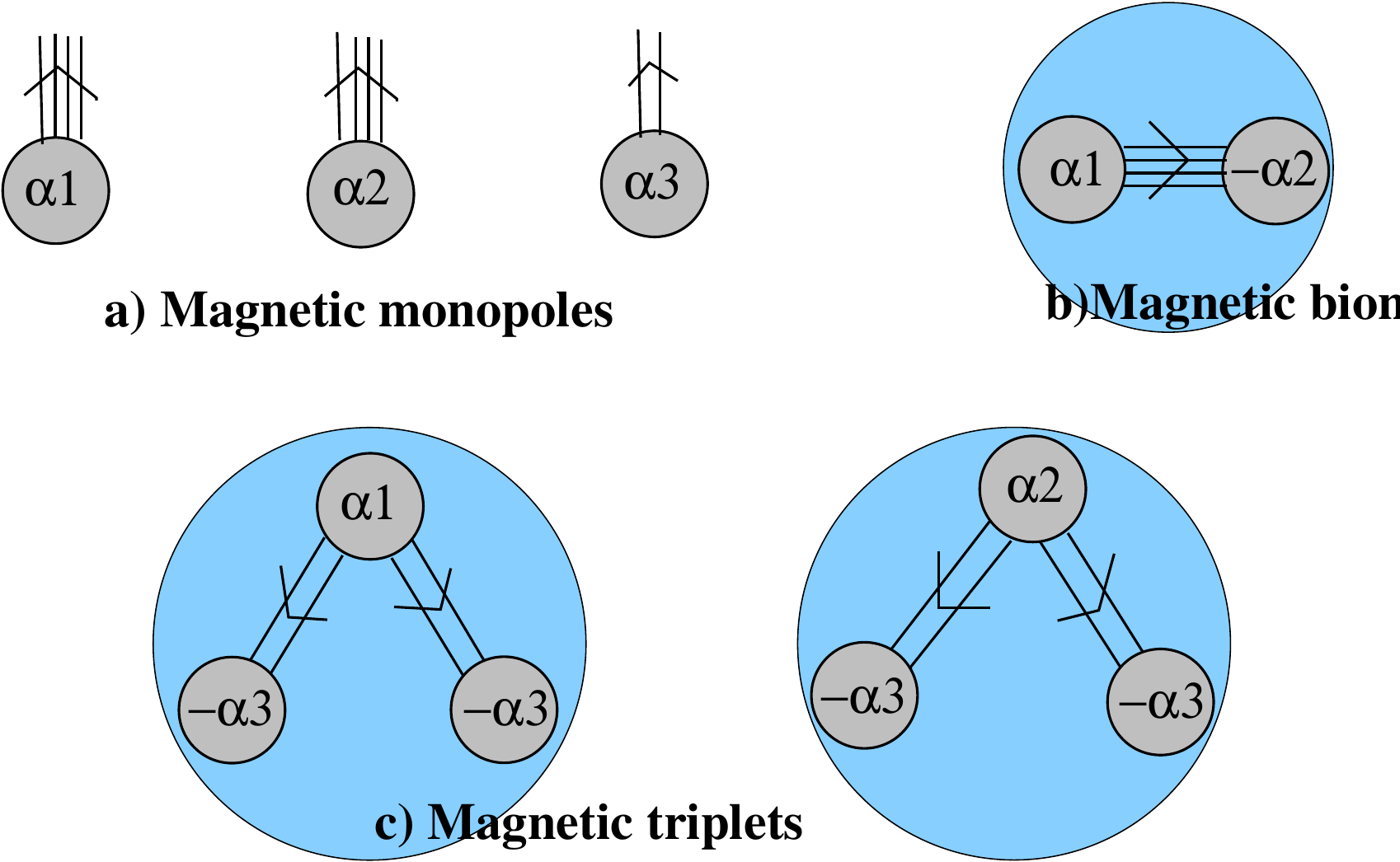}
\caption{
(a)   Monopole operators ${\cal M}_{1}$, ${\cal M}_{2}$, ${\cal M}_{3}$,
with fermionic zero modes dictated by the index theorem,  appearing at order $e^{-S_0}$ in the semiclassical expansion. Monopoles cannot induce confinement due to fermionic zero modes. 
(b)  Magnetic bion ${\cal{B}}_1 \sim {\cal M}_1 { \overline{\cal M}_2} $, which appears at order  $e^{-2S_0}$. 
(c) The two magnetic triplet operators ${\cal T}_1$ and ${\cal T}_2$,   appearing at order $e^{-3S_0}$. A combination of  bions and triplets
leads to a mass gap for the dual photons and confinement in QCD(S).  
 }
  \label {fig:monop}
 \end{center}
 \end{figure}

The invariance of the monopole operators under the $\Z_{10}$ symmetry demands, as usual, 
the dual photons to transform under a $\Z_5$ topological symmetry:
   \begin{eqnarray}
 &&  \psi^4 \longrightarrow  e^{i \frac{8 \pi}{10}}   \psi^4 ,\qquad   e^{i \alpha_1 \cdot \sigma}    \longrightarrow   e^{- i \frac{8 \pi}{10}}   e^{i \alpha_1\cdot \sigma} ~, \cr
 &&  \psi^4 \longrightarrow  e^{i \frac{8 \pi}{10}}   \psi^4 , \qquad   e^{i \alpha_2 \cdot\sigma}    \longrightarrow   e^{- i \frac{8 \pi}{10}}   e^{i \alpha_2\cdot \sigma}  ~, \cr
 &&     \psi^2 \longrightarrow  e^{i \frac{4 \pi}{10}}   \psi^2 ,\qquad   e^{i \alpha_3 \cdot \sigma}    \longrightarrow   e^{- i \frac{4 \pi}{10}}   e^{i \alpha_3 \cdot\sigma}  ~.
     \label{shift2}
   \end{eqnarray}
The  $\Z_5$ shift symmetry forbids all pure flux operators of the type   $e^{-S_0} e^{i \alpha_i \cdot\sigma}$, as  is also clear  from the index theorem.   At order  $e^{-2S_0}$, the only permitted operator is:
   \begin{equation}
 {\cal B}_1 = e^{-2S_0} e^{i (\alpha_1- \alpha_2) \cdot \sigma}~. \label{bionsextet}\end{equation}
  This excitation is  a  magnetic bion, a composite of   
${\cal M}_{\rm 1}   \overline {\cal M}_{\rm 2}$. As there are two types of photons, and the magnetic bion only renders one linear combination massive,  this excitation by itself is not sufficient to generate a mass gap in the gauge sector of QCD(S). At  order  $e^{-3S_0}$, the $\Z_5$ symmetry permits two flux  operators:
 \begin{equation}
  {\cal T}_1 =  e^{-3S_0} e^{i (\alpha_1- 2\alpha_3)\cdot \sigma} \sim 
{\cal M}_{\rm 1}   \overline {\cal M}_{\rm 3}^2, \qquad  {\cal T}_2 = e^{-3S_0} e^{i (\alpha_2- 2\alpha_3)\cdot \sigma}  \sim {\cal M}_{\rm 2}   \overline {\cal M}_{\rm 3}^2
\end{equation}
 which may be referred as  {\it magnetic triplet} operators as they may be viewed as 
 composite of three elementary monopoles, see fig.~\ref{fig:monop}.
 
 Thus, the mass gap in the gauge sector of SU(3) QCD(S) is due to  the magnetic bion and 
triplets $ {\cal B}_1,   {\cal T}_1 ,   {\cal T}_2 $. The  masses of the dual photons can be inferred from the potential terms in the dual lagrangian, which, as (\ref{Eq:dQCD}), is of the form:
 \begin{eqnarray}
 L^3 {\cal L}_{ \rm mass \; gap}  \sim   e^{-2S_0}  \cos (\alpha_1- \alpha_2)\cdot\sigma  + e^{-3S_0}  \cos (\alpha_1- 2\alpha_3)\cdot \sigma + e^{-3S_0}  \cos (\alpha_2- 2\alpha_3)\cdot \sigma ~.
 \label{biontriplet}
\end{eqnarray}
From the one loop definition of the strong scale (\ref{oneloop1}) or (\ref{oneloop2}), we have
 $e^{- S_0} = (\Lambda L)^{b_0}, \; b_0= \beta_0/N= 
\frac{11}{3} - \frac{2}{3} \frac{N+2}{N} N_f^D$. With $N=3, N_f^D=1$,
the masses of the two types of dual photons are proportional to:
\begin{eqnarray}
m_{\sigma_1} \sim \frac{1}{L} e^{-3 S_0/2} = \Lambda (\Lambda L)^{\frac{17}{6} }\; , 
\qquad  m_{\sigma_2} \sim \frac{1}{L} e^{- S_0} =   \Lambda (\Lambda L)^{\frac{14}{9} }~.
\end{eqnarray}

Both mass gaps increase with increasing $L$, suggesting that the SU$(3)$ theory with a single sextet  flavor 
confines on ${\bf R}^4$.  The potential (\ref{biontriplet}) has five isolated vacua in its fundamental domain. The choice of the vacuum completely breaks the ${\mathbb Z}_5$ symmetry, 
leading to the appearance of five  isolated vacua. Since the ${\mathbb Z}_{10}$ chiral symmetry is intertwined with the  ${\mathbb Z}_5$ topological symmetry of the dual photons,  
this breaking generates a 4d (complex) mass term for the fermions.  
 The presence of mass gap and the existence of five isolated vacua induced by chiral symmetry breaking are indeed the expected properties of the QCD(S) on ${\bf R}^4$, providing further evidence for the smoothness conjecture.

\subsection{$\mathbf{SU(N)}$, $\mathbf{N\geq 3}$,  QCD    with $\mathbf{N_f^D}$ two-index symmetric tensor representations. }
\label{suNsymmetric}

The analysis of the generic case with SU(N) gauge group and $N_f^D < N_f^{AF}$ is a combination of the $N_f^D=1$ analysis given above and the analysis of QCD(AS)* given in \S5 of \cite{Shifman:2008ja}. Here, we provide a  summary of the results. 
There are, as usual in SU$(N)$, $N$ types of elementary monopole operators.  The number of fermionic zero modes associated with each of these topological excitations is given by:
\begin{equation}
[ {\cal I}_{1},   {\cal I}_{2}, {\cal I}_{3},   \ldots,  {\cal I}_{N} ] =  N_f^D [ 4,4,2, \ldots, 2], 
 \qquad       {\cal I}_{\rm inst}= \sum_{i=1}^N {\cal I}_{i} = N_f^D (2N+4)
\end{equation}
where   ${\cal I}_{\rm inst}$ is the number of zero modes associated with an instanton. 
Confinement is induced  by $(N-2)$ magnetic bions, which appear at order $e^{-2S_0}$ 
in semiclassical expansion, and by   $2$ magnetic triplets, which appear at order $e^{-3S_0}$. 
The masses for the dual photons are:   
\begin{eqnarray}
m_{\sigma_1} \sim \frac{1}{L} e^{-3 S_0/2} = \Lambda (\Lambda L)^{\frac{9}{2} -  \frac{N+2}{N} N_f^D}  , 
~  \; m_{\sigma_2} \sim  \ldots \sim m_{\sigma_{N-1}}  
\sim \frac{1}{L} e^{- S_0} =   \Lambda (\Lambda L)^{\frac{8}{3} -
 \frac{2}{3} \frac{N+2}{N} N_f^D}.
\end{eqnarray}   According to our conjecture, the theories for which the mass gap vanishes with increasing radius 
 are the ones which flow  to IR-CFTs on ${\bf R}^4$. One subtlety in the case of 
  SU$(N)$ gauge theory with $N_f^D$ symmetric representation fermions is that the mass 
  gap is not solely induced by magnetic bions. The magnetic triplets may also have some impact. 
  But at large $N$, the majority of gauge fluctuations is induced by  the  $N-2$ types of magnetic bions and $2$ types of magnetic triplet operators.  Thus, taking the bion effects as the dominant one, we estimate the conformal window to 
appear in the range: 
  \begin{equation}
  4 \left( 1 - \frac{2}{N+2} \right)  <  N_f^D  < \frac{11}{2}  \left( 1 - \frac{2}{N+2} \right) 
   \qquad    {\rm symmetric, \; vectorlike}~.
 \label{confS}
  \end{equation}
Note that   the $N \rightarrow \infty$ limit of the conformal window for QCD(S)  converges to the same range as the  QCD(adj) with $N_f^W$ adjoint Weyl fermions.  
This is not an accident. In fact, this will be a recurring theme for {\it all} the 
two-index vectorlike {\it and} chiral theories.  The underlying reason is the (nonperturbative) 
large-$N$  orbifold and  orientifold equivalence \cite{Armoni:2004uu,Kovtun:2004bz}.

\subsection{$\mathbf{SU(N)}$ QCD with $\mathbf{N_f^D}$ fundamental fermions}
\label{suNfundamental}

In QCD(F)*, in the small ${\bf S}^1$ regime, there are $N$ types of elementary monopoles. 
The $2N_f$  zero modes of an instanton localize  to one of the monopoles, and the remaining $(N-1)$ monopoles do not carry any fermionic zero modes:
\begin{equation}
[ {\cal I}_{1},   {\cal I}_{2}, {\cal I}_{3},   \ldots,  {\cal I}_{N} ] =  N_f^D [ 2,0, \ldots, 0 ], 
 \qquad       {\cal I}_{\rm inst}=  \sum_{i=1}^N {\cal I}_{i} = 2 N_f^D 
\end{equation}
 Consequently, all the gauge fluctuations acquire mass by monopole operators, all of which appear at order  $e^{-S_0}$.    The characteristic  mass gap, using $b_0=\beta_0/N = 11/3 - 2 N_f^D/(3 N)$, is given by:
\begin{equation}
m_{\sigma} \sim \frac{1}{L} e^{-S_0/2}  =  \Lambda (\Lambda L)^{\frac{b_0}{2}-1} =   \Lambda (\Lambda L)^{ \frac{5 - 2 N_f^D/ N}{6}}~.
\label{massgapfund}
 \end{equation}
  The mass gap is an increasing function of $L$  for  $5 - 2 N_f^D/ N   > 0$, 
 which corresponds to   the confining gauge theories according to our criteria. With increasing $L\Lambda$,  the monopoles become less dilute and the  semiclassical approximation ceases to be valid when   $L \Lambda \sim 1$. 
  The conventional expectation is that such theories  in the   $L \Lambda \sim 1$ regime  must exhibit  non-abelian confinement. 

  For  $5 - 2 N_f^D/ N < 0$, the opposite behavior ensues. The  mass gap is a decreasing function of the radius, which means the characteristic length of gauge fluctuations increases with increasing radius.  Thus, 
  for theories with fundamental fermions, our estimate of the conformal window  is 
  \begin{equation}
  \frac{5}{2}N <  N_f^D  < \frac{11}{2}N, \qquad  {\rm fundamental} \; . 
 \label{confwinF}
  \end{equation}
  
 Note that the two-loop coefficient $\beta_1$ of the QCD(F) theory beta function (see (\ref{beta2loop}))
 flips sign at ${\tilde N_f}= \frac{34N}{13- 3/N^2}$, which
asymptotes
to $2.61N$ at large $N$. Thus, in this class of theories, with  $2.5 N
< N_f < {\tilde N_f}$,   it is strongly plausible that the mass gap
for gauge fluctuations will behave as shown in   fig.~\ref{fig:massgap}c.,
and thus theories with $N_f^*\leq  2.61N$   will exhibit a finite mass gap
for gauge fluctuations in the decompactification limit.
In this case, $N_f=2.5 N$ presents a lower bound on the lower
boundary. The correct values of   $N_f^*$ may as well be slightly
larger than
${\tilde N_f}$.

 \subsection{$\mathbf{SU(N) \times SU(N)}$ QCD with $\mathbf{N_f^D}$ bifundamental fermions}
 \label{suNbifundamental}
 
 QCD(BF) is an SU$(N) \times$SU$(N)$ non-abelian gauge theory with $N_f^D$ Dirac  fermions in the bifundamental representation of the gauge group. 
 For  QCD(BF)* in the  small ${\bf S}^1$ regime,  the structure of the zero modes of monopole operators coincides with QCD(adj) and confinement is induced by magnetic bions which appear at order $e^{-2S_0}$ in topological expansion \cite{Shifman:2008ja}.   Due to simple kinematic reasons, the one-loop beta function of QCD(BF) also coincides with the one of QCD(adj). 
 Explicit computation shows that the 
confining  and conformal range  coincides  with QCD(adj) (modulo the replacement 
$N_f^W \rightarrow N_f^D$)  and is given by: 
 \begin{equation}
  4  <  N_f^D < \frac{11}{2} \qquad  {\rm bifundamental, vectorlike} .
 \label{confwinBF}
  \end{equation}
One may be tempted to think that the matching of the perturbative beta functions of the two theories  is a kinematic accident. However, this is not so.    QCD(BF)  may be obtained by a 
 $\Z_2$ orbifold projection of QCD(adj) with $N_f^W= N_f^D$ adjoint Weyl fermions, and, in the large $N$ limit, there is a non-perturbative equivalence between QCD(adj) and QCD(BF). In fact, 
 a conformal window distinct from (\ref{confwinBF}) (and (\ref{cwa})) would be in  contradiction with the  large-$N$ orbifold equivalence.

  \subsection{$\mathbf{SU(N)}$ QCD with $\mathbf{N_f^D}$ two-index antisymmetric tensor  fermions}
  \label{suNantisymmetric}
  
In QCD(AS)* formulated on  ${\bf R}^3  \times {\bf S}^1$,  the number of fermionic zero modes associated with each one of the N-types of monopoles is given by: 
\begin{equation}
[ {\cal I}_{1},   {\cal I}_{2}, {\cal I}_{3},   \ldots,  {\cal I}_{N} ] =  N_f^D [ 2,2, \ldots, 2, 0,0], 
 \qquad       {\cal I}_{\rm inst}= \sum_{i=1}^N {\cal I}_{i} = N_f^D (2N-4)~,
\end{equation}
where   ${\cal I}_{\rm inst}$ is the number of zero modes associated with an instanton. 
Confinement is induced  by $(N-3)$ magnetic bions, which appear at order $e^{-2S_0}$ 
in the semiclassical expansion, and by  $2$ magnetic monopoles which appear at order $e^{-S_0}$.  
This means that especially the small-$N$ regime of QCD(AS) requires more care. In particular, 
 for SU$(3)$, the range of the conformal window estimated within our approach is same as 
QCD(F) and is given by $7.5 < N_f^D <  16.5$ (of course, this is just because for SU$(3)$, 
QCD(AS)= QCD(F)). 

In general, for SU$(N)$ gauge group, the mass gap for gauge fluctuations in the semiclassical domain is given by:\begin{eqnarray}
m_{\sigma_1} &\sim& m_{\sigma_2} \sim \frac{1}{L} e^{- S_0/2} = \Lambda (\Lambda L)^{\frac{5}{6} -  \frac{1}{3}\frac{N-2}{N} N_f^D}, \nonumber \\
m_{\sigma_3} &\sim&  \ldots \sim m_{\sigma_{N-1}}  
\sim \frac{1}{L} e^{- S_0} =   \Lambda (\Lambda L)^{\frac{8}{3} -
 \frac{2}{3} \frac{N-2}{N} N_f} . 
 \label{gapAS}
\end{eqnarray}
At $N=3$, there is no magnetic bion contribution to the mass gap. Hence, for that case, we must take the estimate coming from the monopole factor, as already explained above. For $N=4$, there are two monopoles and one bion. The magnetic bion enters at a larger length scale.   For this case, we guess that the 
conformal window should start somewhere in between 5 and 8 flavors.
 For $N >  5$, the monopoles are not the major source of confining field configurations. In particular, at large $N$, the monopole contribution is suppressed 
 by $1/N$ relative to magnetic bions. Hence, we expect the conformal window to take place for:  
  \begin{equation}
  4 \left( 1 + \frac{2}{N-2} \right)  <  N_f^D  < \frac{11}{2}  \left( 1 + \frac{2}{N-2} \right) 
   \qquad    {\rm antisymmetric, \; vectorlike} , \; N \ge 5 \; .
 \label{confwinAS}
  \end{equation}

 \section{Conformal windows of some chiral gauge theories}
\label{chiral}
 \smallskip

Our results for the conformal windows of the chiral theories we consider are summarized in Table~\ref{defaultChi}. Notice that the gauge coupling at the fixed-point of the two-loop beta function in the predicted conformal window is 
not large, in most cases. Furthermore, we note that the conformal windows for the chiral theories at large $N$ converge to those for theories with adjoints (see table~\ref{defaultADJ}). It would be of some interest to study the relation of this   to the large-$N$ orbifold/orientifold equivalence; in this regard we note that chiral theories similar to the ones considered here can be obtained by a brane orientifold construction \cite{Lykken:1997ub, Lykken:1998ec}.

  \subsection{Chiral {\bf{SU}}$\mathbf{(N)}$ with $\mathbf{N_f^W}$ generations of ($\mathbf{AS,  \overline F}$)}
 \label{chiralAS}
 
 We first discuss an SU$(N)$ chiral gauge theory with  $N_f^W$  generations of 
 one AS  and $N-4$ anti-fundamental 
 left handed Weyl fermions.  This theory is gauge anomaly free. We wish to determine the conformal window of this class of theories using our techniques. 
 There is compelling evidence \cite{Dimopoulos:1980hn}, based on 't Hooft anomaly matching and complementarity, that one-generation 
 theories on ${\bf{R}}^4$ exhibit confinement without chiral symmetry breaking. 
 On small ${\bf R}^3 \times {\bf{S}}^1$ it was found \cite{Shifman:2008cx}, for $N_f^W=1$, to also exhibit confinement without chiral symmetry breaking and it is expected that  confinement without 
 $\chi$SB holds at any radius and upon decompactification. 
  Hence, the 
  chiral symmetry characterization  and truncated Schwinger-Dyson equations are not (at least naively) a useful tool to study the confinement-conformality transition in this class of theories.  However,  there may be other symmetry-singlet condensates which  are nonvanishing in the confined phase and vanish  in the conformal phase. We are not aware  of such studies regarding chiral theories. 
  
   There are important nonperturbative differences between the chiral and vectorlike theories
with regards to the structure of topological excitations.  Although these will not alter our  simple picture regarding confinement---which is again induced by magnetic bions---it is noteworthy to mention a few. As usual, with the use of the deformation theory, one can make 
the small-${\bf S}^1$ regime accessible to semiclassical analysis.  
  There are $N$ types of monopoles.  The number of fermionic zero modes associated with each of these topological excitation is given by:
 \begin{eqnarray}
        \label{indexchiral}
[ {\cal I}_{1},   {\cal I}_{2}, {\cal I}_{3},   \ldots,  {\cal I}_{N} ] =&&  N_f^W \left(  \underbrace{[ 1,1, \ldots, 1, 0,0] }_{(N-2) AS \;  {\rm zero \;  modes}}
+ \underbrace{ [ 0,0, \ldots, 0, N-4,0]}_ {(N-4) \overline F \;  {\rm zero \;  modes}} \right) ~,\cr \cr 
       {\cal I}_{\rm inst}=&& \sum_{i=1}^N  \; {\cal I}_{i} = N_f^W  [(N-2)AS + (N-4) {\overline F}]~,
 \end{eqnarray}
Distinctly  from vectorlike theories, the generic monopole operators can be  fermionic in this chiral case.
The index (\ref{indexchiral}) is often odd, meaning that the monopole operator is (schematically) of the form:
\begin{eqnarray}
{\cal M}_i &&= e^{-S_0}e^{i \alpha_i \cdot \sigma} \psi_{\rm AS}, \qquad  i=1, \ldots, N-2, \cr 
 {\cal M}_{N-1} &&= e^{-S_0}e^{i \alpha_{N-1} \cdot \sigma} \psi^{i_1}_{\rm \overline F} \ldots  \psi_{\rm \overline F}^{i_{N-4}} \epsilon_{i_1 \ldots i_{N-4}},   \qquad 
  {\cal M}_{N}= e^{-S_0}e^{i \alpha_{N} \cdot \sigma}~.
  \label{suas}
\end{eqnarray}
The $(N-1)$ monopole operators with  fermion zero modes  drop out of dynamics due to averaging over global symmetries, and do not contribute to the nonperturbative dynamics of the theory at order  $e^{-S_0}$.
 In the one-generation model, the first nonperturbatively induced 
global-singlet multi-fermion operator is $\prod_{i=1}^{N-1}{\cal M}_i$, 
which appears at order   $e^{-(N-1)S_0}$.  
Confinement is induced by $(N-2)$-types of magnetic bions and one type of magnetic monopole (${\cal M}_{N}$ in (\ref{suas}), the one whose index (\ref{indexchiral}) is zero). The reader should consult   \cite{Shifman:2008cx} for details about the dynamics of this  theory. 

 The leading-order  beta function  for this theory is:
 \begin{eqnarray}
\beta_0= \frac{11}{3}N  - \frac{2}{3}\left( \frac{N-2}{2} \times 1 + \frac{1}{2} \times 
 (N-4) \right)  N_f^W  = 
 \frac{11}{3}N  - \frac{2}{3} (N-3)N_f^W ~.
\end{eqnarray}
Thus, we consider  asymptotically free SU$(N)$ chiral theories  with:
 \begin{equation}
 N_f^W    [ AS,  (N-4) {\overline F} ] ,  \qquad  N_f^W \leq \frac{11}{2} \left(1 + \frac{3} { N-3} \right) ,  \;   N \geq  5~,
 \end{equation}
 generations. As stated above, the mass gap for gauge fluctuations is induced by magnetic bions 
 (and one magnetic monopole which we neglect below)  and is of order:
\begin{equation}
m_{\sigma} \sim \frac{1}{L} e^{-S_0(L) }\sim    \Lambda (\Lambda L)^{b_0-1}  = \Lambda (\Lambda L)^{ \frac{1}{3}(8 -2 \frac{N- 3}{N} N_f^W)}    ~,
 \end{equation}
 where, as before $b_0 = \beta_0/N$.
Thus, according to our conjecture, the conformal window is expected to appear in the range:
  \begin{equation}
  4 \left( 1 + \frac{3}{N-3} \right)  <  N_f^W  < \frac{11}{2}  \left( 1  +  \frac{3}{N-3} \right) ~,
   \qquad   N_f^W [ {\rm AS},  (N-4) {\overline F} ] , \; N\ge 5.
 \label{confwinASc}
  \end{equation}

\subsection{Chiral {\bf SU}$\mathbf{(N)}$ with  ${\mathbf{N_f^W}}$ generations of ($\mathbf{S,  \overline F}$)}
\label{chiralS}
 Similar consideration also holds for another chiral theory, with 
 $N_f^W$  generations of   $[ S,  (N+4) {\overline F} ] $ chiral 
 matter. The one loop $\beta$-function is:
  \begin{equation}
\beta_0=
 \frac{11}{3}N  - \frac{2}{3} (N+ 3)N_f^W ,    \qquad 1 \leq N_f^W \leq \frac{11}{2} \left(1 - \frac{3} { N+3} \right)  ~.
 \end{equation}
The confinement discussion has similarities with QCD(S) and is dominantly due to magnetic bions (and few magnetic triplets). The bion-induced mass gap is:
\begin{equation}
m_{\sigma} \sim L^{-1} e^{-S_0(L) }\sim    \Lambda (\Lambda L)^{b_0-1}  = \Lambda (\Lambda L)^{ \frac{1}{3}(8 -2 \frac{N+ 3}{N} N_f^W)}    ~.
 \end{equation}
 Thus, the conformal window is expected to appear  in the range:
  \begin{equation}
  4 \left( 1 - \frac{3}{N+3} \right)  <  N_f^W  < \frac{11}{2}  \left( 1  -  \frac{3}{N+3} \right), 
   \qquad   N_f^W[  {\rm S},  (N+4) {\overline F} ] , \; N \geq 5 .
 \label{confwinSc}
  \end{equation}

\subsection{Chiral $\mathbf{SU(N)^K}$ quiver gauge theories} 
\label{chiralquiver}

Chiral quiver gauge theories are gauge theories with  a product gauge group: 
\begin{equation}
 SU(N)_1\times  SU(N)_2 \times \ldots \times  SU(N)_K
\end{equation} 
and chiral  bifundamental  Weyl fermion matter, transforming under the gauge group as:
 \begin{equation}
 \psi_J \sim (1, \ldots, N_J , \overline{N}_{J+1}, \ldots 1), \qquad  J=1, \ldots K, \; \;  
  K+1 \equiv 1 ~.
 \end{equation} 
The nonperturbative aspects of this  class of theories are examined in great detail recently in
\cite{Shifman:2008cx}, and they exhibit many interesting phenomena such as confinement with or without chiral symmetry breaking depending on $N$ and $K$.  We refer the reader to 
\cite{Shifman:2008cx} for detailed discussions. 

 Chiral quiver gauge theories can be obtained by QCD(adj) by orbifold projections. The mechanism of confinement is, as in QCD(adj), 
magnetic bions.  Moreover, the one-loop beta function for each gauge group factor coincides with QCD(adj). Consequently, the conformal window is expected to be in the range: 
  \begin{equation}
  4  <  N_f^W  < \frac{11}{2}, \qquad    {\rm bifundamental, chiral}~.
 \label{confwinBFC}
  \end{equation}

\begin{table}[ht]
\scriptsize
\begin{center}
\begin{tabular}{|c |c |c|c|c|}
\hline
  $N$  & $S+(N+4) {\rm \overline F}$ &  $AS + (N-4) {\rm \overline F}$    &chiral SU$(N)^K$    
  \\ \hline
  5&  $2.5 < N_f^W < 3.43$  &  $10 < N_f^W < 13.75$  &  $4 < N_f^W < 5.5$     \\ \hline
    6 &  $2.67 < N_f^W < 3.67$  &  $8 < N_f^W < 11$  &  $4 < N_f^W < 5.5$   \\ \hline
   7 &   $2.8 < N_f^W < 3.85$ & $7 < N_f^W < 9.63$   &   $4 < N_f^W < 5.5$   \\ \hline
    8&  $2.9 < N_f^W < 4$  & $6.4 < N_f^W < 8.8$   &   $4 < N_f^W <5.5$   \\ \hline
        $\infty$ & $4 < N_f^W < 5.5$   &   $4 < N_f^W < 5.5$ &   $4 < N_f^W < 5.5$    \\ \hline
\end{tabular}
\end{center}
\caption{Estimates  for  the conformal window for  various chiral gauge theories. } 
\label{defaultChi}
\end{table}%

  \section{Comparison with conformal window estimates of other approaches} 
  \label{comparison}
  
  \subsection{Truncated Schwinger-Dyson and NSVZ-inspired beta function studies}
\label{sdcomparison} 

To ease the comparison, we show  the numerical results for the lower boundary of the conformal window for QCD(F/S/AS/Adj) in Tables~\ref{defaultF}, \ref{defaultS}, \ref{defaultAS}, and \ref{defaultADJ}, respectively. We show the values obtained by using our approach (which we refer to as ``deformation theory," or D.T.), the ladder approximation to the truncated Schwinger-Dyson 
equations, and the NSVZ-inspired approach (for which we show  both the $\gamma=2$ and $\gamma=1$ 
results).\footnote{
The values for the ladder approximation to the truncated Schwinger-Dyson equations are taken from 
 ref.~\cite{Dietrich:2006cm}. The two-loop beta function and one-loop anomalous dimension of $\bar\psi \psi$ are used. For QCD(F) we also include  results taken from fig.~3 of ref.~\cite{Gies:2005as}, which  
 uses the functional renormalization group along with the four-loop beta function (the value we give for infinite $N$ is the result of our naive extrapolation of their results from the figure). } The Pad\` e approximation of \cite{Chishtie:1999tx} yields an estimate  consistent with ours: $6\le N_f^{*}\le 9$ for SU$(3)$.
  
  It is interesting to note that for two-index representation fermions, the estimates of our approach and the Schwinger-Dyson equations  differ only by a very small amount 
$| N_f^{*}({\rm ladder}) -  N_f^{*}({\rm D.T.}) | < 0.15$, for $3 \leq N \leq \infty$, whereas they disagree substantially with the $\gamma=2$ NSVZ-inspired estimate  but are close to the latter if $\gamma=1$ is used.

 The situation is reverted for the one-index fundamental representation where $| \frac{N_f^{*}({\rm NSVZ-insp.})}{N} - \frac{ N_f^{*}({\rm D.T.})}{N} | < 0.25$, for $3 \leq N \leq \infty$ and with $\gamma=2$, but the difference with the estimates of the Schwinger-Dyson approach is again quite large (although closer to the functional renormalization group results). The recent observation of \cite{Sannino:2009qc} that a possible magnetic dual to $SU(3)$ QCD(F) loses asymptotic freedom at $N_f > 8$ is also consistent with the estimate from the conjectured beta function with
 $\gamma=2$, as well as with our estimates (which in this case are likely to be lower bounds, see discussion at the end of  section \ref{suNfundamental}).
 
 While we note that the discrepancy is consistent with our estimates being either lower or upper bounds on the boundary of the conformal window,  we would like to argue that, most likely, the estimate of the deformation theory captures the correct regime in both cases. This is because the  deformation theory estimate includes data regarding the mechanism of confinement,  which the other approaches do not see. As shown in Table~\ref{default}, the mass gap and confinement in QCD(F) is due to magnetic monopoles, which enter at order $e^{-S_0}$ in the semiclassical expansion, and for QCD(AS/S/BF/Adj), confinement is  generically due to 
magnetic bions, which appear at order $e^{-2S_0}$ in the topological expansion.  
Below, we will elaborate this statement.

\begin{table}[ht]
\scriptsize
\begin{center}
\begin{tabular}{|c |c |c|c|c|c|c| }
\hline
  $N$  & D.T. (monopoles) \qquad & Ladder (SD)-approx.& Functional RG & NSVZ-inspired: $\gamma = 2$/$\gamma = 1$  & $N_f^*\big\vert_{\beta_1}$ & $N_f^{AF}$ \\ \hline
  2 & $ 5 $ &    $7.85 $  & $8.25$  & $5.5$/7.33& 5.55& 11        \\ \hline
  3 & $7.5 $ &   $11.91  $  &  $10$& $8.25$/11 & 8.05& 16.5   \\ \hline
    4 &  $10 $          & $15.93$ &  $13.5$ &$11$/14.66 &10.61 & 22           \\ \hline
   5 &   $12.5 $  &   $19.95  $   &$16.25$  &$13.75$/18.33 & 13.2 & 27.5  \\ \hline
    10 &   $25 $    &  $39.97 $  & n/a &$ 27.5$/36.66 &26.2 & 55  \\ \hline
        $\infty$ & $ 2.5 N$  &  $  4 N$& $\sim (2.75-3.25)N$&$ 2.75N$/$3.66N$   & 2.61$N$& $5.5N$ \\ \hline
\end{tabular}
\end{center}
\caption{Estimates  for lower boundary of conformal window for  QCD(F), $N_f^* < N_f^D  < 5.5 N$.  To support the discussion in the text, see section 4.3, we have  also given the number of flavors where the two-loop coefficient of the beta function flips sign, $N_f^*\big\vert_{\beta_1}$.} 
\label{defaultF}
\end{table}%
\begin{table}[ht]
\scriptsize
\begin{center}
\begin{tabular}{|c |c |c|c|c|}
\hline
  $N$  & D.T. (bions) & Ladder (SD)-approx. & NSVZ-inspired: $\gamma = 2$/$\gamma = 1$  & $N_f^{AF}$
  \\ \hline
  3 & 2.40  & 2.50   & 1.65/2.2 & 3.30 \\ \hline
    4 & 2.66 & 2.78   & 1.83/2.44 & 3.66 \\ \hline
   5 & 2.85 & 2.97  & 1.96/2.62  & 3.92 \\ \hline
    10 & 3.33 & 3.47  & 2.29/3.05 &4.58 \\ \hline
        $\infty$ & 4  & 4.15 & 2.75/3.66 &5.5  \\ \hline
\end{tabular}
\end{center}
\caption{ Estimates for lower boundary of  conformal window   in QCD(S), $N_f^* < N_f^D < 5.5  \left( 1 - \frac{2}{N+2} \right) $. }
\label{defaultS}
\end{table}%
\begin{table}[ht]
\scriptsize
\begin{center}
\begin{tabular}{|c |c |c|c|c|}
\hline
  $N$  & D.T. (bions) &  Ladder (SD)-approx. & NSVZ-inspired: $\gamma = 2$/$\gamma = 1$   &  $N_f^{AF}$  \\ \hline
    4 & 8   & 8.10   & 5.50/7.33 & 11 \\ \hline
   5 & 6.66  & 6.80 & 4.58/6.00 & 9.16  \\ \hline
    6 & 6  & 6.15  & 4.12/5.5 &  8.25 \\ \hline
    10 & 5& 5.15 & 3.43/4.58 &  6.87 \\ \hline
        $\infty$ & 4  & 4.15 & 2.75/3.66 & 5.50 \\ \hline
\end{tabular}
\end{center}
\caption{ Estimates for lower boundary of  conformal window   in QCD(AS), $N_f^* < N_f^D < 5.5  \left( 1 + \frac{2}{N-2} \right)$. }
\label{defaultAS}
\end{table}%
 \begin{table}[ht]
\scriptsize
\begin{center}
\begin{tabular}{|c |c |c|c|c|}
\hline
  $N$  & D.T. (bions) & Ladder (SD)-approx. & NSVZ-inspired: $\gamma = 2$/ $\gamma = 1$  & $N_f^{AF}$ \\ \hline
  any $N$ & 4    & 4.15   & 2.75/3.66 & 5.5 \\ \hline
\end{tabular}
\end{center}
\caption{ Estimates for lower boundary of  conformal window   in QCD(adj), $N_f^* < N_f^W < 5.5  $. In QCD(adj), we count the number of Weyl fermions as opposed to Dirac, since the adjoint representation is real.}
\label{defaultADJ}
\end{table}%

 \subsection{Kinematics, dynamics, and  universality}
\label{universality}

The beta functions of   QCD-like theories depend on the matter content and may be viewed as  
 kinematic data, counting degrees of freedom. 
For the purpose of computing the beta function, when $N$ is sufficiently large, at leading order we have 
the following relations between different representations of the fermionic matter:  
\begin{eqnarray}
{\rm  \; adjoint \;  Weyl  }&=&   {\rm AS\;  Dirac =  S\;  Dirac= } \; N \times { \rm  (  F \;   Dirac) } \cr
&=& 
[{\rm   AS\;  Weyl },   N \times ({\rm \overline F \;   Weyl }) ] 
= 
[{\rm    S\;  Weyl },   N \times ({\rm \overline F \;   Weyl }) ] ~,
\end{eqnarray}
in the sense that 
$N_f$ multiples of any of the above will induce the same beta function, at leading order in $N$, as  can be checked explicitly.  Note that in this limit one adjoint Weyl fermion counts as one Dirac AS/S, and since the adjoint representation is real, we count it in Weyl-fermion multiples.  
We can define  the parameter:
\begin{equation}
\xi^{AF} ({\cal R})= \left\{ \frac{N_f^{AF, D} ({\rm F})} {N}, \;  N_f^{AF,D}( {\rm AS/S/BF}),    \;  N_f^{AF, {\rm W}}(\rm Adj)\right\}~, 
\label{afparameter}
\end{equation}
characterizing the asymptotic freedom boundary for the representation $\cal{R}$.
It is not hard to see that, for all gauge theories examined in this paper, including the chiral ones, the upper boundary of the conformal window converges to a universal number regardless of what theory we deal with:
\begin{equation}
\lim_{ N \rightarrow \infty} \xi^{AF} ({\cal R}) =5.5
\end{equation}
There is no dynamics entering to this universal number, it follows from counting. 

We can also define a similar parameter for the lower boundary of conformal window. This parameter 
is crucial to isolate kinematic effects from dynamic effects in various approximations. It is defined similar to (\ref{afparameter}), by replacing $N_f^{AF}$ by $N_f^*$: 
\begin{equation}
\xi^{*} ({\cal R})= \left\{ \frac{N_f^{*, D} ({\rm F})} {N}, \;  N_f^{*,D}({\rm AS/S/BF}),    \;  N_f^{*, {\rm W}}(\rm Adj)\right\} ~.
\end{equation}
Now, let us find the $\xi^{*} ({\cal R})$ ratio for the various analytic estimates  for the conformal window boundary:
\begin{eqnarray}
& {\rm Ladder\; (or \;functional \;RG)} :&   \lim_{ N \rightarrow \infty} \xi^{*} ( {\rm F}) = 4 \;  ({\rm or}\sim 3) , \qquad  \lim_{ N \rightarrow \infty} \xi^{*} (\rm S/AS/BF/Adj) =4.15~({\rm n/a}), \cr \cr
& {\rm NSVZ-inspired}, \gamma=2:  & \lim_{ N \rightarrow \infty} \xi^{*} ( {\rm F}) =  \lim_{ N \rightarrow \infty} \xi^{*} (\rm S/AS/BF/Adj) =2.75 ~,\cr \cr
& {\rm NSVZ-inspired}, \gamma=1:  & \lim_{ N \rightarrow \infty} \xi^{*} ( {\rm F}) =  \lim_{ N \rightarrow \infty} \xi^{*} (\rm S/AS/BF/Adj) =3.66 ~,\cr \cr
& {\rm Deformation\;  theory}:  & \lim_{ N \rightarrow \infty} \xi^{*} ( {\rm F}) = 2.5, \qquad 
 \lim_{ N \rightarrow \infty} \xi^{*} (\rm S/AS/BF/Adj/chiral) =4 ~.
\end{eqnarray}

The fact that the  parameter  $\xi^{*} ({\cal R})$ is  different in our estimates is tied with the different mechanism of confinement operating in one- or two-index theories, as stated earlier.  With monopole- and bion-induced confinements, 
the increasing or decreasing behavior of the mass gap as  a function of $L$ is determined 
by:
\begin{equation}
{\rm sign}\left( \frac{b_0}{2}-1\right)  \qquad {\rm and } \qquad {\rm sign}\left( b_0-1\right)~,
\end{equation}
respectively, where $b_0 = \beta_0/N$. If the sign is positive, the mass gap increases upon approaching $\bf{R}^4$, while for negative sign, it decreases. This dynamical information---monopoles vs. bions as the mechanism generating the mass gap---is not present in the other analytic approaches. It is the difference between the monopole- and bion-induced confinement that accounts for the different predictions of the deformation theory approach. At the same time, it should be noted that our approach shares a quality common with the other analytic approaches---the errors of our estimates are hard to evaluate.

 \subsection{Lattice gauge theory studies}
 \label{lattice}
 
In recent years, many groups have begun lattice studies aiming to first identify the conformal window in various vectorlike theories (and, later, to study the properties of the (nearly-)CFTs). Below, we summarize recent lattice results on the conformality vs. confinement issue and compare them with our bounds. We find that these are consistent with our estimates of  the conformal window (remembering their rough nature). In various cases, different lattice studies claim results not consistent with each other, indicating that more work is required to obtain more definite results. However, we expect that in the next few years the lattice results in this regard will become more precise.
 
{\flushleft{Q}CD(F): }
 
The most studied example is that of SU$(3)$ gauge theories with a varying number of fundamental Dirac flavors, $N_f^D$.
 The study of \cite{Iwasaki:2003de}, using Wilson fermions, argued that the conformal window in SU$(3)$ QCD(F) is $7 \le N_f^D \le 16$. 
 Their estimate disagrees with the recent studies  \cite{Appelquist:2007hu, Appelquist:2009ty} of a gauge invariant nonperturbatively defined running coupling  (via the Schr\" odinger functional  using exactly massless staggered fermions)  arguing that $N_f^D=8$ lies outside the conformal window, while $N_f^D = 12$ is conformal.
 Thus, \cite{Appelquist:2007hu, Appelquist:2009ty} place the  lower boundary  in the interval $8 < N_f^* < 12$. 
 
 The results of  \cite{Iwasaki:2003de} also disagree with the study \cite{Deuzeman:2008sc} of the $N_f^D = 8$ theory 
claiming evidence for a true continuum (rather than a lattice-artifact) first-order thermal phase transition between a chirally-symmetric and chirally-broken phase  (using staggered fermions at one value of the mass, argued to be sufficiently small \cite{Deuzeman:2008sc}). Such a transition is not expected to occur in the continuum chirally symmetric CFT  at finite temperature, hence the authors argued that the eight-flavor theory is confining.
An analysis by the same group \cite{Deuzeman:2009mh}  of the $N_f^D=12$ theory  found  that the location of the transition there is insensitive to the physical temperature and is thus a ``bulk" transition to the strongly-coupled confining and chirally-broken phase, a lattice artifact, implying thus consistency with a continuum CFT behavior. 
 
On the other end of the spectrum (as far as estimates for the QCD(F) conformal window) ref.~\cite{Fodor:2008hn}  measured  the low-lying eigenvalues of the staggered Dirac operator, and argued that both $N_f^D =8,12$ are consistent with confinement and chiral symmetry  breaking. At the same time, the authors stated that more studies of taste-breaking artifacts are needed to reach a definite conclusion.

These estimates, save for the study of \cite{Iwasaki:2003de} (the only one consistent with deformation theory taken at face value), 
 are in (rough) agreement with the data given in Table~\ref{defaultF}, given the uncertain nature of the theoretical estimates. 

  {\flushleft{Q}CD(S): }
  
The dynamics of SU$(3)$ with two flavors ($N_f^D=2$) of sextet Dirac fermions was recently studied by several groups \cite{Shamir:2008pb,Fodor:2008hm,DeGrand:2008kx}.
Our estimate of the lower boundary of the SU$(3)$ theory with Dirac sextets is $N_f^D = 2.4$ (see Table~\ref{defaultS})  and taken at face value would imply that the two-flavor theory is confining.

In \cite{Shamir:2008pb} a zero of the discrete beta function defined by the Schr\" odinger functional  on rather small lattices was found, while the further study \cite{DeGrand:2008kx} of the finite-$T$ confinement-deconfinement transition,   using Wilson fermions, argued for consistency with the existence of an IR fixed point. The study of \cite{Fodor:2008hm} is the only one using chiral dynamical quarks  also claimed possible consistency with an IR fixed point, but was not conclusive due to the small volumes and statistics. See also the recent study of \cite{DeGrand:2009et} of the volume scaling of the lowest eigenvalues of the Dirac operator.

  {\flushleft{Q}CD(adj): }
 
 SU$(2)$ with $N_f^W=4$ adjoint Weyl fermions (2 Dirac flavors) has been the subject of the recent lattice studies of \cite{Catterall:2007yx,DelDebbio:2008zf,Catterall:2008qk,Hietanen:2008mr,Hietanen:2009az}. The results of \cite{Catterall:2007yx, Catterall:2008qk} are consistent with either a conformal behavior or ``walking" behavior and more studies are needed to be more conclusive. While the studies  of \cite{Hietanen:2008mr,DelDebbio:2008zf}
are similarly inconclusive, the more recent study \cite{Hietanen:2009az} of the running gauge coupling defined via the Schr\" odinger functional finds evidence for conformal behavior. This is consistent with our estimate and those of other approaches,   given in Table~\ref{defaultADJ}. Recall also, see \S\ref{suNsymmetric}, that the accuracy of our estimate does not allow us to be definite about $N_f^W = 4$. 

\section{Conclusions and prospects}
\label{conclusions} 

We proposed  a new method to determine the infrared behavior of asymptotically free 
Yang-Mills theories with massless vectorlike and chiral matter content. This technique and our estimates  differs from all other existing methods in the literature in various ways.  While all other existing methods are strongly  influenced by two-loop perturbation theory, this data never enters  our non-perturbative semiclassical analysis. Thus, there is no a priori  reason for our approach to produce estimates  in the same domain as  other approaches, 
as we saw in \S\ref{comparison}.

 The data crucial  for the estimates of our approach is the 
 knowledge of  the mechanism of confinement in the semiclassical regime of these gauge theories. This knowledge was gathered over the last two years (confinement is due to magnetic monopoles, bions, triplets, quintets, depending on the particulars of the theory) using the calculability of the  mass gap for gauge fluctuations 
 in some domain of a circle compactification on  ${\bf R}^3  \times {\bf S}^1$.  We introduced a mass gap criterion to distinguish the conformal theories from confining theories on   ${\bf R}^4$. This characterization  also usefully applies to chiral gauge theories, for which a gauge invariant fermion bilinear does not exist. We 
 conjectured that the behavior of mass gap as a function of radius, as shown  in fig.~\ref{fig:massgap},  provides a characterization of conformal versus confining theories.

 As emphasized in \S\ref{non-selfduality}, what makes the composite topological excitations 
 so elusive is their non-(anti)selfduality. This means that they do not arise as solutions of  Prasad-Sommerfield-type equations.  Nonetheless, they  are dynamically  stable due to a fermionic ``pairing" mechanism, 
 and carry non-vanishing magnetic charge. Their action 
 is a fraction of the 4d instanton action, $\frac{2}{N}S_{\rm inst}$ for magnetic bions  and  $\frac{3}{N}S_{\rm inst}$ for magnetic triplets.

In the future, it may be possible to improve our estimates by calculating the so-far-ignored prefactors of $g$ in magnetic bion and triplet operators.  This will generate  $\log (\Lambda L)$ corrections in our mass gap formulas, and may be useful in determining various cases of indeterminacy.

\begin{figure}[t]
\centering
\includegraphics[width=5in]{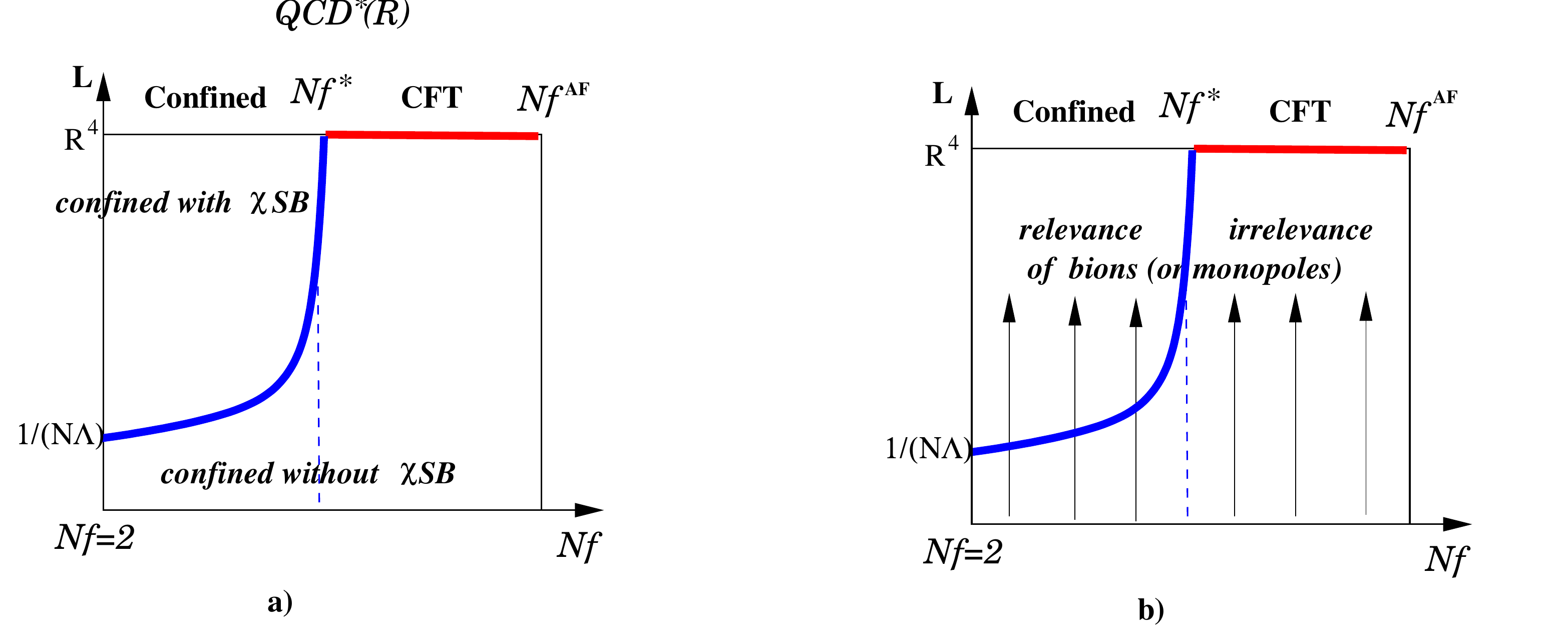}
\caption{
Phase diagram of the center stabilized QCD*(F) theories as a function of
$N_f$ and $L$. Shown are only $N_f \ge 2$ theories with a continuous chiral
symmetry, which may exhibit a $\chi$SB transition as a function of $L$. {\bf a.)} QCD* theories with $N_f< N_f^*$ exhibit  confinement with and without
chiral symmetry breaking as a function of radius. QCD* theories with $N_f>
N_f^*$ and fixed point  at weak coupling   exhibit confinement  without $\chi$SB at any
finite radius  (however, at large $L$, the onset scale of confinement
is so large that any foreseeable simulations will identify this phase
as abelian Coulomb phase). At $L=\infty$, all QCD* theories with $N_f>
N_f^*$ flow to a CFT. 
{\bf b.)} depicts the   main idea of the paper that the mass gap induced by
topological excitations
becomes IR relevant or irrelevant for the two class of theories.
The dashed center line is to guide to eye, there is no transition
there and everything is smooth.
$1/(N\Lambda)$ is the natural scale (somehow counter-intuitively)  of
chiral symmetry breaking.
}
 \label {fig:concon}
\end{figure}

An  alternative approach, similar to the one of this paper  and expected to work for theories with non-trivial center, such as QCD(adj), QCD(BF), and chiral quiver theories, is to compactify these theories on a small ${\bf T}^3 \times {\bf R}$ and impose twisted boundary conditions on gauge fields and fermions, by generalizing  the analysis of \cite{GonzalezArroyo:1995zy}. 
Presumably, at  arbitrary weak coupling, twisting will be sufficient to stabilize the center symmetry. In such a case, the deformations are not needed and small 
 and large volume theories may indeed be smoothly connected in the sense of center symmetry. 
It may be interesting to see if one can probe the roots of conformality or confinement upon compactification to quantum mechanics or other toroidal geometries, 
${\bf T^d} \times {\bf R}^{4-d}$, with $d=2,3,4.$   In particular, making progress on ${\bf T}^4$ by using either deformation theory or twisted boundary conditions may provide direct comparison with lattice gauge theory.  

  Finally, we note that recently  it was conjectured that for QCD(F) in the large $N$ limit, the transition from the chiral symmetry broken confined phase to the chirally symmetric conformal phase as a function of 
 $N_f/N$  on ${\bf R}^4$  may be of the Berezinsky-Kosterlitz-Thouless-type (BKT). The conjecture is based on some 
 similarity between the RG description of the BKT transition and properties of the $\chi$SB-conformal transition in QCD(F), see ref.~\cite{Kaplan:2009kr} for details. 
 
 The physics behind the canonical BKT-transition is that the topological excitations (there: $XY$-model vortices) are relevant in one phase and irrelevant in the other. 
 This is reminiscent of the dilution versus non-dilution of our topological excitations as a function of ${\bf S}^1$ radius. However, on a small $ {\bf S}^1 \times {\bf R}^3$, there is really no phase transition between theories with $N_f \leq N_f^*$ and  $N_f > N_f^*$. This is because at small $L$ both classes of theories exhibit confinement {\it  without } $\chi$SB, albeit  the first class with a mass gap for gauge fluctuations increasing with the radius and the latter with a gap, which decreases with the radius. Thus, in the small-$L$ regime, there is no sign of a BKT transition, and the two 
 regimes are smoothly connected as a function of  $N_f/N$. However, if $L$ is sufficiently large, the  $N_f \leq N_f^*$ theories are expected to exhibit confinement with $\chi$SB.  On the other hand, at  $L$ large  the deformed theories QCD(F)$^*$  with $N_f > N_f^*$ flow to CFTs and there is no $\chi$SB.
 Thus, at sufficiently large ${\bf S^1}$,  a critical line exists between these two phases as shown on figure~\ref{fig:concon},  possibly of a BKT-type.  The  relevance versus irrelevance of   topological excitations is probably necessary, but perhaps not sufficient to show that this transition is  BKT-like.     It would be interesting  to examine the relation between the BKT-conjecture of  \cite{Kaplan:2009kr} and our proposal in more detail.

 \acknowledgments We  thank O. Aharony, J. Giedt, B. Holdom, K. Intriligator, D.B. Kaplan, B. Mueller, T. Okui, M. Peskin, T. Ryttov, F. Sannino,   S. Shenker, M. Shifman, and J. Terning for useful discussions.  
 This work was supported by the U.S.\ Department of Energy Grant DE-AC02-76SF00515 and by the National Science and Engineering Council of Canada (NSERC).

\appendix

\section{Conventions:  $\beta$-functions and strong scale}
\label{conventions}

The scheme-independent two loop beta function for $G=SU(N)$ gauge theory with $N_f$ 
Dirac fermions in representation $R$ of the gauge group $G$ is given by: 
\begin{eqnarray}
\label{beta2loop}
&&\frac{\partial g}{\partial {\rm log} \mu} =  \beta(g) = -\frac{\beta_0}{(4 \pi)^2} g^3 -  \frac{\beta_1}{(4 \pi)^4} g^5  ~,\cr \cr
&&\beta_0= \frac{11}{3}C_2(G) - \frac{4}{3} T(R) N_f ~,\cr \cr
&& \beta_1= \frac{34}{3} C_2(G)^2 - \frac{20}{3} C_2(G) T(R)N_f  - 4 C_2 (R) T(R) N_f  ~.
\end{eqnarray}
Here, $C_2(R)$ is the quadratic Casimir for  representation $R$, $C_2(G)$---the quadratic Casimir for adjoint  representation, $d(R)$ is dimension of $R$,  $d(G)$ is dimension of the group $G$, and 
$T(R) \delta^{ab} = \tr T^a_{R} T^b_{R}$ is the trace normalization for representation $R$:
\begin{equation}
C_2(R) d(R) = d(G)T(R) ~.
\end{equation}
 For $SU(N)$, 
 \begin{eqnarray}
&& C_2(G)=N, \qquad  d(G)= N^2-1,   \cr \cr 
&& T(R) =\left \{\frac{1}{2}, \frac{N+2}{2}, \frac{N-2}{2}, N \right\} \qquad {\rm for} \; R= \{ \rm F, S, AS, Adj\}   \cr \cr
&& d(R) = \left \{N, \frac{N(N+1)}{2}, \frac{N(N-1)}{2}, N^2-1 \right \} \cr \cr
&& C_2(R) =  \left \{\frac{N^2-1}{2N}, \frac{(N-1)(N+2)}{N}, \frac{(N+1)(N-2)}{N}, N \right \}
\end{eqnarray}
The asymptotic freedom boundary is: 
\begin{eqnarray}
N_f^{AF} [R] &&= \frac{11}{4} \frac{d(G) C_2(G)}{d(R) C_2(R)} = \frac{11}{4}   \frac{C_2(G)}{T(R)} 
\cr \cr 
 && =     \frac{11}{4} \left \{2N,  \frac{ 2N}{N+2},   \frac{ 2N}{N-2},   
 1 \right \}
 \label{upper}
 \end{eqnarray}  
We define the strong scale by using  one-loop beta function as , 
\begin{equation}
\Lambda^{\beta_0} = \mu^{\beta_0} e^{- \frac{8 
\pi^2}{g^2({\mu})}}~.
\label{oneloop1}
\end{equation}
One should note that  $\frac{8 \pi^2}{g^2({\mu})}= S_{\rm inst}$ is the usual BPST
 instanton action.  One should also note that the usual instanton effects are of order 
 $e^{-S_{\rm inst}} \sim e^{-N } $ and are suppressed in the 't Hooft's large $N$ limit, with $g^2 N$ fixed.   In the study of center symmetric (or approximately center symmetric gauge theories on  
 $ {\bf R}^3 \times  {\bf S}^1$), the semiclassical expansion is an expansion 
 in   $e^{- \frac{8 \pi^2}{g^2{(L)}N}} = e^{-S_0}$, where $S_0$ is the action of BPS or KK monopoles. These objects generically carry fractional topological charge of $1/N$ (the BPST 
 instanton charge is normalized to one).  Thus   we rewrite the one loop result  (\ref{oneloop1})  as: 
 \begin{equation}
\Lambda^{b_0} = \mu^{b_0} e^{- \frac{8 
\pi^2}{g^2{(\mu)}N}}, \qquad  b_0 \equiv \frac{\beta_0}{N}, \qquad {\rm equivalently} \qquad 
e^{-S_0(L)}= (\Lambda L)^{b_0}~,
\label{oneloop2}
\end{equation}
 where in the final formula, we used $\mu = \frac{1}{L}$.


\bigskip 

\begin{thebibliography}{99}
  
\bibitem{Holdom:1981rm}
  B.~Holdom,
  ``Raising The Sideways Scale,''
  Phys.\ Rev.\  D {\bf 24}, 1441 (1981).
 
\bibitem{Yamawaki:1985zg}
  K.~Yamawaki, M.~Bando and K.~i.~Matumoto,
  ``Scale Invariant Technicolor Model And A Technidilaton,''
  Phys.\ Rev.\ Lett.\  {\bf 56}, 1335 (1986).

\bibitem{Appelquist:1986an}
  T.~W.~Appelquist, D.~Karabali and L.~C.~R.~Wijewardhana,
   ``Chiral Hierarchies and the Flavor Changing Neutral Current Problem in
  Technicolor,''
  Phys.\ Rev.\ Lett.\  {\bf 57}, 957 (1986).


  
\bibitem{Luty:2004ye}
  M.~A.~Luty and T.~Okui,
  ``Conformal technicolor,''
  JHEP {\bf 0609}, 070 (2006)
  [arXiv:hep-ph/0409274].
 
\bibitem{Georgi:2007ek}
  H.~Georgi,
  ``Unparticle Physics,''
  Phys.\ Rev.\ Lett.\  {\bf 98}, 221601 (2007)
  [arXiv:hep-ph/0703260].
  
  
\bibitem{Randall:1999ee}
  L.~Randall and R.~Sundrum,
 ``A large mass hierarchy from a small extra dimension,''
  Phys.\ Rev.\ Lett.\  {\bf 83}, 3370 (1999)
  [arXiv:hep-ph/9905221].
  
\bibitem{ArkaniHamed:2000ds}
  N.~Arkani-Hamed, M.~Porrati and L.~Randall,
  ``Holography and phenomenology,''
  JHEP {\bf 0108}, 017 (2001)
  [arXiv:hep-th/0012148].
  
\bibitem{Rattazzi:2008pe}
  R.~Rattazzi, V.~S.~Rychkov, E.~Tonni and A.~Vichi,
  ``Bounding scalar operator dimensions in 4D CFT,''
  JHEP {\bf 0812}, 031 (2008)
  [arXiv:0807.0004 [hep-th]].
 
\bibitem{Rychkov:2009ij}
  V.~S.~Rychkov and A.~Vichi,
 ``Universal Constraints on Conformal Operator Dimensions,''
  arXiv:0905.2211 [hep-th].
  
\bibitem{Banks:1981nn}
  T.~Banks and A.~Zaks,
   ``On The Phase Structure Of Vector-Like Gauge Theories With Massless
  Fermions,''
  Nucl.\ Phys.\  B {\bf 196}, 189 (1982).
 
 
\bibitem{Peskin:1982mu}
  M.~E.~Peskin,
``Chiral Symmetry And Chiral Symmetry Breaking,'' Lectures presented at the Summer School on Recent Developments in Quantum Field Theory and Statistical Mechanics, Les Houches, France, Aug 2 - Sep 10, 1982 (North Holland, 1984). 

  
\bibitem{Appelquist:1988yc}
  T.~Appelquist, K.~D.~Lane and U.~Mahanta,
``On the ladder approximation for spontaneous chiral symmetry breaking,"
  Phys.\ Rev.\ Lett.\  {\bf 61}, 1553 (1988).
  
\bibitem{Cohen:1988sq}
  A.~G.~Cohen and H.~Georgi,
  ``Walking Beyond The Rainbow,''
  Nucl.\ Phys.\  B {\bf 314}, 7 (1989).

\bibitem{Miransky:1996pd}
  V.~A.~Miransky and K.~Yamawaki,
  ``Conformal phase transition in gauge theories,''
  Phys.\ Rev.\  D {\bf 55}, 5051 (1997)
  [Erratum-ibid.\  D {\bf 56}, 3768 (1997)]
  [arXiv:hep-th/9611142].
  
\bibitem{Appelquist:1996dq}
  T.~Appelquist, J.~Terning and L.~C.~R.~Wijewardhana,
  ``The Zero Temperature Chiral Phase Transition in SU(N) Gauge Theories,''
  Phys.\ Rev.\ Lett.\  {\bf 77}, 1214 (1996)
  [arXiv:hep-ph/9602385].

\bibitem{Appelquist:1998rb}
  T.~Appelquist, A.~Ratnaweera, J.~Terning and L.~C.~R.~Wijewardhana,
  ``The phase structure of an SU(N) gauge theory with N(f) flavors,''
  Phys.\ Rev.\  D {\bf 58}, 105017 (1998)
  [arXiv:hep-ph/9806472].
  
\bibitem{Dietrich:2006cm}
  D.~D.~Dietrich and F.~Sannino,
 ``Walking in the SU(N),''
  Phys.\ Rev.\  D {\bf 75}, 085018 (2007)
  [arXiv:hep-ph/0611341].
 
\bibitem{Gies:2005as}
  H.~Gies and J.~Jaeckel,
  ``Chiral phase structure of QCD with many flavors,''
  Eur.\ Phys.\ J.\  C {\bf 46}, 433 (2006)
  [arXiv:hep-ph/0507171].

\bibitem{Appelquist:1997dc}
  T.~Appelquist and S.~B.~Selipsky,
 ``Instantons and the chiral phase transition,''
  Phys.\ Lett.\  B {\bf 400}, 364 (1997)
  [arXiv:hep-ph/9702404].
 
\bibitem{Velkovsky:1997fe}
  M.~Velkovsky and E.~V.~Shuryak,
``QCD with large number of quarks: Effects of the instanton  anti-instanton
  pairs,''
  Phys.\ Lett.\  B {\bf 437}, 398 (1998)
  [arXiv:hep-ph/9703345].
 
 
 
\bibitem{Dimopoulos:1980hn}
  S.~Dimopoulos, S.~Raby and L.~Susskind,
   ``Light Composite Fermions,''
  Nucl.\ Phys.\  B {\bf 173}, 208 (1980).


 
\bibitem{Intriligator:1995au}
  K.~A.~Intriligator and N.~Seiberg,
``Lectures on supersymmetric gauge theories and electric-magnetic  duality,''
  Nucl.\ Phys.\ Proc.\ Suppl.\  {\bf 45BC}, 1 (1996)
  [arXiv:hep-th/9509066].
 
   
\bibitem{Intriligator:2005if}
  K.~A.~Intriligator,
  ``IR free or interacting? A proposed diagnostic,''
  Nucl.\ Phys.\  B {\bf 730}, 239 (2005)
  [arXiv:hep-th/0509085].

\bibitem{Novikov:1983uc}
  V.~A.~Novikov, M.~A.~Shifman, A.~I.~Vainshtein and V.~I.~Zakharov,
   ``Exact Gell-Mann-Low Function Of Supersymmetric Yang-Mills Theories From
  Instanton Calculus,''
  Nucl.\ Phys.\  B {\bf 229}, 381 (1983).
 
\bibitem{Novikov:1985rd}
  V.~A.~Novikov, M.~A.~Shifman, A.~I.~Vainshtein and V.~I.~Zakharov,
  ``Beta Function In Supersymmetric Gauge Theories: Instantons Versus
  Traditional Approach,''
  Phys.\ Lett.\  B {\bf 166}, 329 (1986)
  [Sov.\ J.\ Nucl.\ Phys.\  {\bf 43}, 294.1986\ YAFIA,43,459 (1986\ YAFIA,43,459-464.1986)].

  
\bibitem{Ryttov:2007cx}
  T.~A.~Ryttov and F.~Sannino,
  ``Supersymmetry Inspired QCD Beta Function,''
  Phys.\ Rev.\  D {\bf 78}, 065001 (2008)
  [arXiv:0711.3745 [hep-th]].

\bibitem{Grinstein:2008qk}
  B.~Grinstein, K.~A.~Intriligator and I.~Z.~Rothstein,
  ``Comments on Unparticles,''
  Phys.\ Lett.\  B {\bf 662}, 367 (2008)
  [arXiv:0801.1140 [hep-ph]].

\bibitem{Ryttov:2007sr}
  T.~A.~Ryttov and F.~Sannino,
   ``Conformal Windows of SU(N) Gauge Theories, Higher Dimensional
  Representations and The Size of The Unparticle World,''
  Phys.\ Rev.\  D {\bf 76}, 105004 (2007)
  [arXiv:0707.3166 [hep-th]].
  
\bibitem{Chishtie:1999tx}
  F.~A.~Chishtie, V.~Elias, V.~A.~Miransky and T.~G.~Steele,
 ``Pade summation approach to QCD beta-function infrared properties,''
  Prog.\ Theor.\ Phys.\  {\bf 104}, 603 (2000)
  [arXiv:hep-ph/9905291].
  
 
\bibitem{Appelquist:1999hr}
  T.~Appelquist, A.~G.~Cohen and M.~Schmaltz,
  ``A new constraint on strongly coupled field theories,''
  Phys.\ Rev.\  D {\bf 60}, 045003 (1999)
  [arXiv:hep-th/9901109].
  
\bibitem{Poppitz:2009kz}
  E.~Poppitz and M.~\" Unsal,
  ``Chiral gauge dynamics and dynamical supersymmetry breaking,''
  arXiv:0905.0634 [hep-th].
  
  

\bibitem{Shifman:1999mv}
  M.~A.~Shifman and A.~I.~Vainshtein,
  ``Instantons versus supersymmetry: Fifteen years later,''
  arXiv:hep-th/9902018.


\bibitem{Sannino:2005sk}
  F.~Sannino,
  ``Higher representations: Confinement and large N,''
  Phys.\ Rev.\  D {\bf 72}, 125006 (2005)
  [arXiv:hep-th/0507251].


\bibitem{Unsal:2007jx}
  M.~\" Unsal,
  ``Magnetic bion condensation: A new mechanism of confinement and mass gap in
  four dimensions,''
  arXiv:0709.3269 [hep-th].
  
\bibitem{Shifman:2008ja}
  M.~Shifman and M.~\" Unsal,
  ``QCD-like Theories on $R_3\times S_1$: a Smooth Journey from Small to Large
  $r(S_1)$ with Double-Trace Deformations,''
  Phys.\ Rev.\  D {\bf 78}, 065004 (2008)
  [arXiv:0802.1232 [hep-th]].
  
\bibitem{Unsal:2008ch}
  M.~\" Unsal and L.~G.~Yaffe,
  ``Center-stabilized Yang-Mills theory: confinement and large $N$ volume
  independence,''
  Phys.\ Rev.\  D {\bf 78}, 065035 (2008)
  [arXiv:0803.0344 [hep-th]].

  
\bibitem{Shifman:2008cx}
  M.~Shifman and M.~\" Unsal,
  ``On Yang-Mills Theories with Chiral Matter at Strong Coupling,''
  arXiv:0808.2485 [hep-th].
  
\bibitem{Shifman:2009tp}
  M.~Shifman and M.~\" Unsal,
 ``Multiflavor QCD$^*$ on $R^3 \times S^1$: studying transition from abelian to
  non-abelian confinement,''
  arXiv:0901.3743 [hep-th].   
  
 
\bibitem{Seiberg:1996nz}
  N.~Seiberg and E.~Witten,
`` Gauge dynamics and compactification to three dimensions,"
  hep-th/9607163.
  
\bibitem{Aharony:1997bx}
  O.~Aharony, A.~Hanany, K.~A.~Intriligator, N.~Seiberg and M.~J.~Strassler,
  ``Aspects of N = 2 supersymmetric gauge theories in three dimensions,''
  Nucl.\ Phys.\  B {\bf 499}, 67 (1997)
  [arXiv:hep-th/9703110].
     
      \bibitem{Davies:1999uw}
  N.~M.~Davies, T.~J.~Hollowood, V.~V.~Khoze and M.~P.~Mattis,
   ``Gluino condensate and magnetic monopoles in supersymmetric  gluodynamics,''
  Nucl.\ Phys.\  B {\bf 559}, 123 (1999)
  [arXiv:hep-th/9905015].
 
\bibitem{Myers:2007vc}
  J.~C.~Myers and M.~C.~Ogilvie,
  ``New Phases of SU(3) and SU(4) at Finite Temperature,''
  Phys.\ Rev.\  D {\bf 77}, 125030 (2008)
  [arXiv:0707.1869 [hep-lat]].
  
\bibitem{Ogilvie:2007tj}
  M.~C.~Ogilvie, P.~N.~Meisinger and J.~C.~Myers,
  ``Exploring Partially Confined Phases,''
  PoS {\bf LAT2007}, 213 (2007)
  [arXiv:0710.0649 [hep-lat]].

\bibitem{Myers:2008zm}
  J.~C.~Myers and M.~C.~Ogilvie,
  ``Exotic phases of finite temperature SU(N) gauge theories,''
  Nucl.\ Phys.\  A {\bf 820}, 187C (2009)
  [arXiv:0810.2266 [hep-th]].
  
 
\bibitem{Cossu:2009sq}
  G.~Cossu and M.~D'Elia,
  ``Finite size phase transitions in QCD with adjoint fermions,''
  arXiv:0904.1353 [hep-lat].

\bibitem{Meisinger:2009ne}
  P.~N.~Meisinger and M.~C.~Ogilvie,
  ``String Tension Scaling in High-Temperature Confined SU(N) Gauge Theories,''
  arXiv:0905.3577 [hep-lat].
 



\bibitem{Nye:2000eg}
  T.~M.~W.~Nye and M.~A.~Singer,
  ``An $L^2$-Index Theorem for Dirac Operators on $S^1 \times R^3$,''
  arXiv:math/0009144.

   
\bibitem{Poppitz:2008hr}
  E.~Poppitz and M.~\" Unsal,
  ``Index theorem for topological excitations on $R^3 \times S^1$ and Chern-Simons
  theory,''
  JHEP {\bf 0903}, 027 (2009)
  [arXiv:0812.2085 [hep-th]].


\bibitem{Lee:1997vp}
  K.~M.~Lee and P.~Yi, ``Monopoles and instantons on partially compactified D-branes,
Phys.\ Rev.\  D {\bf 56}, 3711 (1997), 
  arXiv:hep-th/9702107.

\bibitem{Kraan:1998pm}
  T.~C.~Kraan and P.~van Baal, ``Periodic instantons with non-trivial holonomy,"
 Nucl.\ Phys. B {\bf 533}, 627 (1998), arXiv:hep-th/9805168

\bibitem{Armoni:2004uu}
 A.~Armoni, M.~Shifman and G.~Veneziano,
 ``From super-Yang-Mills theory to QCD: Planar equivalence and its
 implications,''
 arXiv:hep-th/0403071.



\bibitem{Kovtun:2004bz}
 P.~Kovtun, M.~Unsal and L.~G.~Yaffe,
  ``Necessary and sufficient conditions for non-perturbative equivalences  of
 large N(c) orbifold gauge theories,''
 JHEP {\bf 0507}, 008 (2005)
 [arXiv:hep-th/0411177].

\bibitem{Lykken:1997ub}
  J.~D.~Lykken, E.~Poppitz and S.~P.~Trivedi,
``M(ore) on chiral gauge theories from D-branes,''
  Nucl.\ Phys.\  B {\bf 520}, 51 (1998)
  [arXiv:hep-th/9712193].
\bibitem{Lykken:1998ec}
  J.~D.~Lykken, E.~Poppitz and S.~P.~Trivedi,
  ``Branes with GUTs and supersymmetry breaking,''
  Nucl.\ Phys.\  B {\bf 543}, 105 (1999)
  [arXiv:hep-th/9806080].

\bibitem{Iwasaki:2003de}
  Y.~Iwasaki, K.~Kanaya, S.~Kaya, S.~Sakai and T.~Yoshie,
``Phase structure of lattice QCD for general number of flavors,''
  Phys.\ Rev.\  D {\bf 69}, 014507 (2004)
  [arXiv:hep-lat/0309159].
     
\bibitem{Appelquist:2007hu}
  T.~Appelquist, G.~T.~Fleming and E.~T.~Neil,
  ``Lattice Study of the Conformal Window in QCD-like Theories,''
  Phys.\ Rev.\ Lett.\  {\bf 100}, 171607 (2008)
  [arXiv:0712.0609 [hep-ph]].
  
\bibitem{Appelquist:2009ty}
  T.~Appelquist, G.~T.~Fleming and E.~T.~Neil,
  ``Lattice Study of Conformal Behavior in SU(3) Yang-Mills Theories,''
  arXiv:0901.3766 [hep-ph].
    
    
\bibitem{Deuzeman:2008sc}
  A.~Deuzeman, M.~P.~Lombardo and E.~Pallante,
  ``The physics of eight flavours,''
  Phys.\ Lett.\  B {\bf 670}, 41 (2008)
  [arXiv:0804.2905 [hep-lat]].


\bibitem{Deuzeman:2009mh}
  A.~Deuzeman, M.~P.~Lombardo and E.~Pallante,
``Evidence for a conformal phase in SU(N) gauge theories,''
  arXiv:0904.4662 [hep-ph].
 

\bibitem{Fodor:2008hn}
  Z.~Fodor, K.~Holland, J.~Kuti, D.~Nogradi and C.~Schroeder,
   ``Probing technicolor theories with staggered fermions,''
  arXiv:0809.4890 [hep-lat].
 
\bibitem{Shamir:2008pb}
  Y.~Shamir, B.~Svetitsky and T.~DeGrand,
   ``Zero of the discrete beta function in SU(3) lattice gauge theory with color
  sextet fermions,''
  Phys.\ Rev.\  D {\bf 78}, 031502 (2008)
  [arXiv:0803.1707 [hep-lat]].
 
\bibitem{DeGrand:2008kx}
  T.~DeGrand, Y.~Shamir and B.~Svetitsky,
  ``Phase structure of SU(3) gauge theory with two flavors of
 symmetric-representation fermions,''
  Phys.\ Rev.\  D {\bf 79}, 034501 (2009)
  [arXiv:0812.1427 [hep-lat]].
 
 
\bibitem{Fodor:2008hm}
  Z.~Fodor, K.~Holland, J.~Kuti, D.~Nogradi and C.~Schroeder,
   ``Nearly conformal electroweak sector with chiral fermions,''
  arXiv:0809.4888 [hep-lat].


 
\bibitem{DeGrand:2009et}
  T.~DeGrand,
 ``Volume scaling of Dirac eigenvalues in SU(3) lattice gauge theory with
   color sextet fermions,''
  arXiv:0906.4543 [hep-lat].
 
   
     
\bibitem{Catterall:2007yx}
  S.~Catterall and F.~Sannino,
   ``Minimal walking on the lattice,''
  Phys.\ Rev.\  D {\bf 76}, 034504 (2007)
  [arXiv:0705.1664 [hep-lat]].

\bibitem{Catterall:2008qk}
  S.~Catterall, J.~Giedt, F.~Sannino and J.~Schneible,
   ``Phase diagram of SU(2) with 2 flavors of dynamical adjoint quarks,''
  JHEP {\bf 0811}, 009 (2008)
  [arXiv:0807.0792 [hep-lat]].
  
\bibitem{Hietanen:2008mr}
  A.~J.~Hietanen, J.~Rantaharju, K.~Rummukainen and K.~Tuominen,
  ``Spectrum of SU(2) lattice gauge theory with two adjoint Dirac flavours,''
  JHEP {\bf 0905}, 025 (2009)
  [arXiv:0812.1467 [hep-lat]].

\bibitem{DelDebbio:2008zf}
  L.~Del Debbio, A.~Patella and C.~Pica,
   ``Higher representations on the lattice: numerical simulations. SU(2) with
  adjoint fermions,''
  arXiv:0805.2058 [hep-lat].
    
\bibitem{Hietanen:2009az}
  A.~J.~Hietanen, K.~Rummukainen and K.~Tuominen,
  ``Evolution of the coupling constant in SU(2) lattice gauge theory with two
 adjoint fermions,''
  arXiv:0904.0864 [hep-lat].


\bibitem{Sannino:2009qc}
  F.~Sannino,
  ``QCD Dual,''
  arXiv:0907.1364 [hep-th].

\bibitem{GonzalezArroyo:1995zy}
  A.~Gonzalez-Arroyo and P.~Martinez,
  ``Investigating Yang-Mills theory and confinement as a function of the
  spatial volume,''
  Nucl.\ Phys.\  B {\bf 459}, 337 (1996)
  [arXiv:hep-lat/9507001].


\bibitem{Kaplan:2009kr}
  D.~B.~Kaplan, J.~W.~Lee, D.~T.~Son and M.~A.~Stephanov,
  ``Conformality Lost,''
  arXiv:0905.4752 [hep-th].
   
  
\end{thebibliography}
\end{document}